# Dynamics of Null and Electrostatic Blind Spots for Quantitative PFM

Marti Checa[1*], Holland Swicegood[1], Ruben Millan-Solsona[1], Ralph Bulanadi[1], Jerry Zhao[1], Jack Lasseter[1], Stephen Jesse[1], Liam Collins[1*]

1 Center for Nanophase Materials Sciences, Oak Ridge National Laboratory, Oak Ridge, Tennessee 37831, USA.

email: CHECAM@ORNL.GOV, COLLINSLF@ORNL.GOV

Piezoresponse force microscopy (PFM) is a cornerstone technique for probing nanoscale electromechanical phenomena, yet quantitative—and in some cases qualitative—interpretation remains hindered by parasitic electrostatic forces coupled through cantilever dynamics. Recent approaches aim to suppress these artifacts by operating at resonance-defined null spots (NS) or electrostatic blind spots (ESBS), but whether these conditions are equivalent—and how they evolve under realistic measurement conditions—has remained unclear. Here, combining analytical beam models, geometrically faithful finite-element simulations, and automated interferometric measurements, we show that NS and ESBS are fundamentally different operating conditions. The NS is a modal zero at contact resonance where sensitivity to all excitations vanishes, whereas the ESBS is a quasistatic position where only the distributed electrostatic response is suppressed. Automated measurements reveal that both conditions evolve with the tip–sample boundary condition yet remain spatially separated under realistic experimental conditions. While beam models capture the dominant cantilever mechanics, finite-element simulations and experiment show that quantitative prediction of near-tip behavior requires realistic three-dimensional probe electrostatics and mechanics. These findings establish NS and ESBS as dynamic operating conditions and provide a practical framework for advancing interferometric PFM toward truly quantitative electromechanical metrology.

## Introduction

Piezoresponse Force Microscopy (PFM) is an atomic force microscopy (AFM)-based technique for imaging and quantifying nanoscale electromechanical coupling in piezoelectric and ferroelectric materials[1,2]. A conductive AFM tip is placed in contact with the sample while an AC voltage applied between the tip and the bottom electrode (or sample) induces local surface strain via the converse piezoelectric effect[3,4]. The resulting cantilever oscillation is detected—typically by optical beam deflection (OBD) or interferometric displacement sensing (IDS)—to map the effective piezoelectric coefficient ($d_{eff}$) and polarization orientation[5]. The same detection principles underpin electrochemical strain microscopy (ESM)[6], which instead probes bias-induced Vegard strain arising from ionic redistribution, extending these measurements to mixed ionic–electronic conductors and energy materials[7-9].

Despite its ubiquity, quantitative interpretation remains challenging. The measured signal is a convolution of the local electromechanical response with long-range electrostatic forces filtered through cantilever dynamics. In particular, both distributed electrostatic forces acting along the cantilever and localized tip–sample interactions drive cantilever motion, producing responses that can be difficult to distinguish from a genuine piezoresponse. The resulting transfer function depends sensitively on probe geometry, tip–sample boundary conditions (e.g., contact stiffness and damping), excitation frequency, and detection position, giving rise to spatially and spectrally varying artifacts. Consequently, the detected signal reflects not only the true surface displacement at the tip–sample junction but the coupled electromechanical response of the cantilever–sample system. Long-range capacitive forces can further induce cantilever bending independent of local tip displacement, generating measurable signals even in the absence of intrinsic electromechanical activity. [10-12] Together, these effects can produce image contrast and ferroelectric-like hysteresis loops in nominally non-piezoelectric materials. [13,14] As a result, the detection modality—specifically how and where cantilever motion is measured—plays a central role in determining measurement fidelity.

To mitigate these issues, recent techniques target specific sensing locations: IDS can operate at resonant null spots (NS) near the cantilever tip, while slope-based OBD targets electrostatic blind spots (ESBS) near the cantilever mid-section. It has remained unclear, however, whether these operating conditions—NS and ESBS—are equivalent or precisely how stable they are under realistic experimental conditions. To resolve these questions, we combine one and two-segment analytical beam models (Methods; Supplementary Notes 1 and 2), automated interferometric PFM measurements, and geometrically faithful finite-element simulations. The analytical beam models identify the governing cantilever mechanics and distinguish NS from ESBS. Experiments track both conditions as a function of applied load, and finite-element simulations help pinpoint the additional three-dimensional probe physics required for quantitative agreement. Collectively, these approaches reveal that both slope- and displacement-based detection exhibit distinct NS and ESBS that remain spatially separated under realistic measurement conditions, with their positions evolving with the tip–sample boundary condition and realistic near-tip probe geometry. Because these operating conditions evolve throughout an experiment, positioning the detection laser at a presumed null or blind spot cannot reliably suppress electrostatic artifacts. The analytical beam models capture the dominant cantilever mechanics and predict the qualitative evolution of the NS and ESBS; however, quantitative prediction of the near-tip mode shapes—and the finite separation that persists between the D-NS and D-ESBS at high load—requires realistic three-dimensional probe mechanics, including tip compliance and distributed electrostatic loading, which are accurately recovered only by the finite-element model. These findings have direct implications for quantitative imaging and nanoscale hysteresis measurements, particularly in materials with weak electromechanical response where micrometer-scale sensing-position errors can generate picometer-scale electrostatic crosstalk comparable to the intrinsic signal.

## Slope and displacement null spots:

Most commercial AFMs employ OBD, in which a laser reflected from the cantilever onto a split photodiode measures cantilever angular deflection. Consequently, OBD is sensitive to cantilever slope ($\partial z/\partial x$) rather than absolute tip displacement. Interferometric approaches, including laser Doppler vibrometry and quadrature phase differential interferometry,[5,15] instead recover the out-of-plane displacement at a defined position along the cantilever.[10] This distinction is critical in PFM because localized electromechanical motion beneath the tip and distributed body-electrostatic forces excite different beam shapes[10,16,17]; as noted above, the latter can generate an apparent piezoresponse even in non-piezoelectric materials. These distinct spatial responses create detection-specific positions at which the body-electrostatic contribution is suppressed while sensitivity to tip-localized motion is retained. Such positions have been identified for both detection modalities: an interior electrostatic blind spot for slope-sensitive OBD and a position near the tip for displacement-sensitive interferometry. [5,10,18-21] These special sensing locations have enabled improved rejection of body electrostatic artifacts and more direct measurement of sample displacement; interferometric implementations have further supported quantitative, vector-resolved reconstruction of polarization responses.[18,20] Beyond establishing their existence and practical utility, however, the physics governing these locations remains incompletely understood. In particular, it remains unclear whether the null positions observed under contact-resonance operation[5] are equivalent to the electrostatic blind spots identified under quasi-static conditions[22], or how their locations depend on contact stiffness, probe geometry, and cantilever dynamics.

To address these questions, **Fig. 1** summarizes the contact-resonance Euler–Bernoulli model of Bradler *et al.*,[23] which incorporates finite tip height and mass, cantilever tilt and damping, normal and longitudinal contact mechanics, and both localized and distributed excitation pathways (**Fig. 1a**). Because the tip of a real probe is recessed from the cantilever free end, we extend this formulation to a two-segment beam in which the tip–sample reaction acts at an interior contact point while the short distal overhang terminates at a true free end (Methods; Supplementary Note 1). **Figure 1** illustrates the transition from distinct quasistatic responses to a common contact-resonance mode shape. Under quasistatic excitation ($f \ll f_{CR}$; **Fig. 1b,d**), localized piezoelectric motion and distributed body-electrostatic forcing generate different spatial responses, producing detector-specific electrostatic blind spots where the electrostatic contribution vanishes while sensitivity to tip-localized motion is retained. At contact resonance ($f = f_{CR}$; **Fig. 1c,e**), both excitation pathways project onto the same contact-resonance eigenmode, causing the normalized displacement and slope profiles to collapse onto one another. The resulting null spots are therefore modal zeros rather than electrostatic blind spots: at these locations, the measured slope or displacement vanishes, simultaneously suppressing both electromechanical and electrostatic contributions (**Fig. S1**).

The framework developed here connects two experimental strategies that evolved largely independently. When IDS was introduced for quantitative PFM, the interferometer was positioned directly above the tip by iteratively minimizing the resonant transfer function. [5] This near-tip location was treated as a displacement null intended to suppress cantilever-dynamic contributions while retaining direct sensitivity to tip motion. The ESBS concept emerged separately for OBD, where the laser is positioned to null the slope response to distributed body electrostatics while preserving sensitivity to localized electromechanical deformation.[22] Experimentally, the ESBS is identified by minimizing the bias-dependent response on a non-ferroelectric reference or by satisfying the equal-amplitude condition on oppositely polarized ferroelectric domains.[22] These approaches therefore came to be viewed as parallel artifact-mitigation strategies, despite being defined in different frequency regimes and by different physical criteria.

More recent subresonant comparisons reported an IDS "null point" reproducibly at or near the tip ($x/L \approx 1$), whereas the corresponding OBD position occupied a broad, condition-dependent region of the cantilever.[19] Within the framework established here, however, these subresonant positions correspond to the D-ESBS and S-ESBS, respectively, rather than to resonant null spots. The apparent stability of the near-tip IDS position should also be interpreted in light of the probe "seasoning" performed before measurement, which preconditions the tip–sample junction and may preferentially retain probes and contact states that respond reproducibly near the tip. Although this provides a useful reference condition, comparable stability is not guaranteed when load, contact stiffness, and electrostatic coupling evolve during general PFM imaging or spectroscopy.[19] The four detector-specific operating conditions are summarized in **Table 1**.

Interpreting these near-tip displacement conditions requires a model that represents the geometry beyond the tip–sample contact. Earlier forced-response models treated the cantilever as a single beam terminating at the contact and represented the remaining overhang through an effective tip mass.[19,23] This approximation accurately reproduces the response over most of the cantilever, but its validity has not been established for operating points located at or beyond the contact. Our formulation instead follows Rabe's treatment of the tip as an interior contact while explicitly retaining the distal overhang and its true free-end boundary condition.[24,25] **Figure 1** uses this two-segment formulation to distinguish the resonant null spots from the subresonant electrostatic blind spots for both slope- and displacement-based detection.

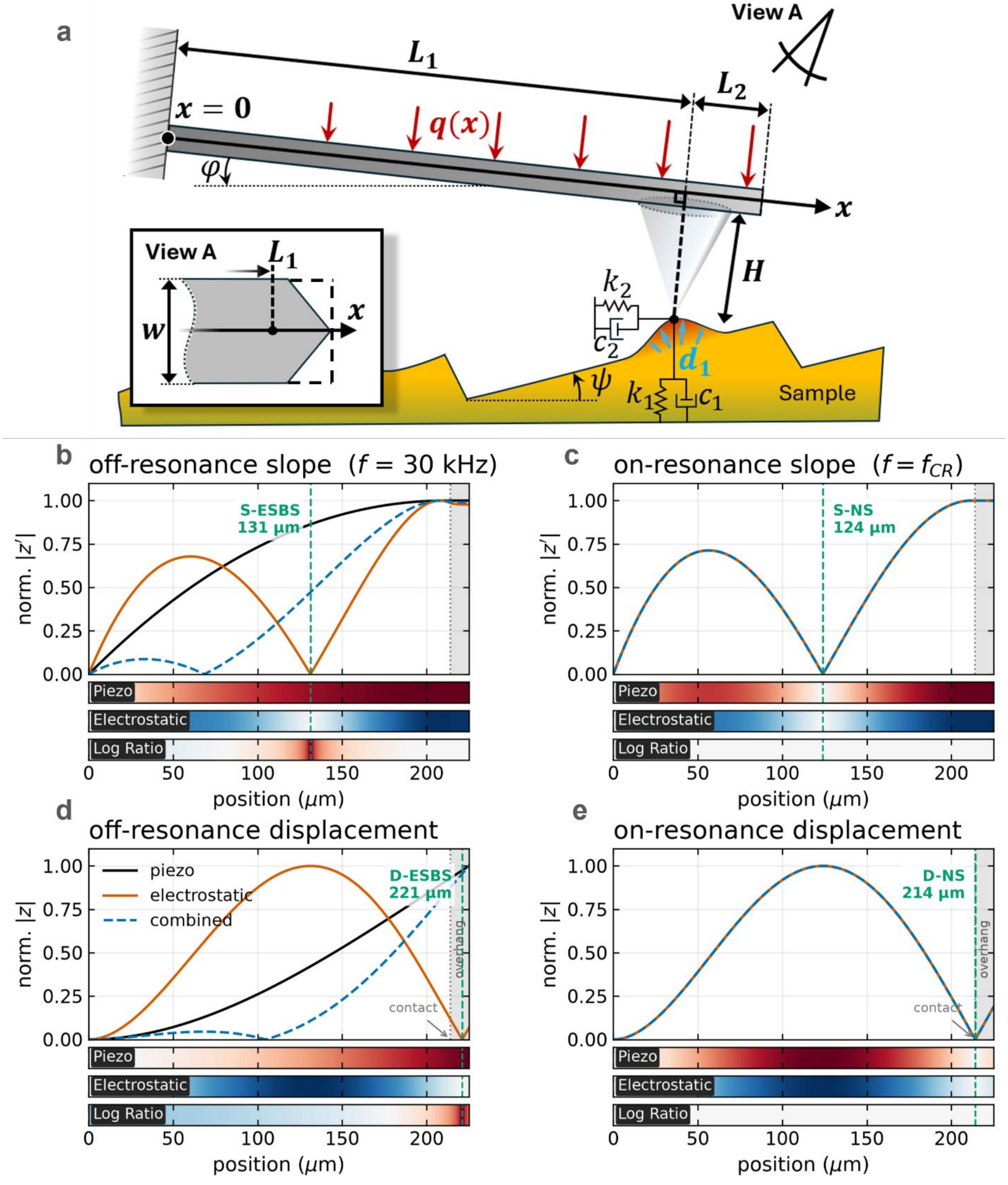

**Figure 1. | Euler–Bernoulli model of the cantilever response under piezoelectric and electrostatic excitation.** (a) Schematic of the two-segment Euler–Bernoulli (EB) cantilever model, adapted from Bradler et al., showing a clamped AFM cantilever of length L, width w, and tip height H under a distributed electrostatic force q(x). The tip–sample interaction is represented by normal and lateral contact stiffness and damping elements ($k_1$, $k_2$ and $c_1$, $c_2$), with sample-surface inclination ψ and tip displacement ztip. Because the tip of real probes is set back

from the free end, the model is extended to two segments: the tip reaction acts at an interior contact point $x_c = L - \Delta x$, and the remaining overhang carries a true free end (Methods; Supplementary Note 1). Panels (b–e) are computed for a BudgetSensors Multi75E-G probe (Table S1) in the frictionless limit ($k_2 = 0$) at $k_1 = 1500$ N/m; the effective compliance-limited $k_2$ that captures the elastic tip cone is treated separately in Supplementary Note 1. In each of (b–e), the vertical dotted line marks the interior contact point and the grey band the free overhang beyond it. (b) Off-resonance slope response ($\partial z/\partial x$) versus detection position for purely piezoelectric (black), purely electrostatic (red), and combined (blue dashed) excitation; the lower color maps show the corresponding signal magnitude. The green dashed line marks the slope electrostatic blind spot (S-ESBS). (c) On-resonance slope response for the first in-contact bending mode; the green dashed line marks the slope null spot (S-NS). (d) Off-resonance displacement response (z); the green dashed line marks the displacement electrostatic blind spot (D-ESBS). (e) On-resonance displacement response; the green dashed line marks the displacement null spot (D-NS). The slope conditions (b,c) lie near mid-lever, whereas the displacement blind spot and null spot (d,e) fall on the overhang and at the contact point, respectively — the near-tip region that exists only in the two-segment model. Comparison of (b–e) shows that the electrostatic blind spot and the null spot are distinct conditions that generally occur at different detection positions along the cantilever.

The slope operating points in panels (**b,c**) lie near the middle of the cantilever, whereas the displacement operating points in panels (**d,e**) lie within the final ~5% of the cantilever length beyond the contact. With the tip–sample contact located at $x/L = 0.952$, the D-NS occurs immediately beyond the contact while the D-ESBS lies farther along the distal overhang toward the free end. Their prediction therefore depends explicitly on the mechanics, distributed electrostatic loading, and true free-end boundary condition of this post-contact segment. A single-segment beam terminating at the contact cannot fully represent this region.[19]

Comparison with the original formulation confirms that the overhang has negligible influence over most of the cantilever (**Fig. S2**). Its effect is confined primarily to the near-tip region in which the displacement operating points reside—the region central to the present work. Figure 1 also shows that the electrostatic blind spots and resonant null spots are physically distinct conditions that generally occur at different detection positions. **Figure 2** therefore examines how these positions evolve with contact stiffness before comparison with automated interferometric measurements and geometrically faithful finite-element simulations.

| **Condition** | **Definition** | **Frequency regime** | **Implication for PFM operation** |
|---|---|---|---|
| S-NS<br>(slope null spot) | Detection position where the resonant slope response (OBD) vanishes | At contact resonance<br>(sub or intra resonance nulls also possible) | Not a valid measurement position —<br>electromechanical and electrostatic responses are suppressed together; migrates strongly (tens of μm) with load and contact stiffness |

| D-NS (displacement null spot) | Detection position where the resonant displacement response (IDS) vanishes | At contact resonance (sub or intra resonance nulls also possible) | Not a valid measurement position — sensitivity to all excitations is lost; lies within a few µm of the tip, migrates with load; in the beam model its high-load limit is set by the lateral contact stiffness and tip geometry |
|---|---|---|---|
| S-ESBS (slope electrostatic blind spot) | Detection position where the slope response to the distributed (body) electrostatic force vanishes while the local electromechanical response remains finite | Quasistatic ($f \ll f_{CR}$) | Valid operating point for OBD-PFM; lies near the cantilever mid-point and exists at essentially any real contact, but drifts with contact stiffness. |
| D-ESBS (displacement electrostatic blind spot) | Detection position where the displacement response to the distributed (body) electrostatic force vanishes while the local electromechanical response remains finite | Quasistatic ($f \ll f_{CR}$) | Valid operating point for IDS-PFM; lies near the tip but exists only above a probe-dependent contact stiffness, drifts with load and near-tip electrostatic loading (tip height, tilt, taper) |

**Table 1.** Summary and definitions of null spots (NS) and electrostatic blind spots (ESBS) for slope- and displacement-based detection.

## Null Spots and Electrostatic Blind Spots as distinct conditions:

Whereas **Fig. 1** establishes the conceptual distinction between null spots and electrostatic blind spots, **Fig. 2** shows how their positions evolve with the tip–sample boundary condition. The frequency–position maps in **Fig. 2a** display the complete slope and displacement transfer functions as the contact stiffness (k*) is increased. Resonances appear as amplitude ridges, while transfer-function zeros appear as minima accompanied by phase reversals. A resonant slope or displacement NS occurs where the corresponding zero intersects the first contact-resonance branch. At the lowest contact stiffness, a displacement null spot is absent whereas the slope null spot is already present. As the contact stiffens and the effective boundary condition evolves from compliant toward pinned, the displacement null emerges and both operating points migrate toward the cantilever interior. The migration is substantially larger for the S-NS, although the D-NS also shifts measurably.

To assess the importance of increasingly realistic physical assumptions, we benchmarked four beam-level contact models spanning the conventional single-segment Euler–Bernoulli formulation of Bradler et al.[19] through a cubic-spline overhang approximation, an explicit two-segment overhang model, and finally a two-segment model incorporating finite tip-apex compliance. This hierarchy isolates the effects of overhang geometry and tip compliance on contact resonances, mode shapes, and near-tip null positions, and identifies which predicted observables are most sensitive to the underlying physics (Fig. S16). The maps in **Fig. 2** use an elastic-tip contact, in which the isotropic tip–sample springs act through the compliant silicon tip cone (lateral stiffness $k_{cone}$); this is the one-dimensional analogue of the three-dimensional finite-element contact and, unlike a rigid lateral spring, preserves the pinned-like near-tip mode shape (Supplementary Note 1). **Figure 2b,c** compares the D-NS with the D-ESBS on the same spatial coordinate. The D-NS is identified from the resonant displacement zero, where sensitivity to both localized and distributed excitation is lost, whereas the D-ESBS is identified off resonance from the zero of the distributed electrostatic response, where the localized electromechanical response remains finite. The two conditions must therefore be determined in different frequency regimes. The D-NS and D-ESBS remain distinct across compliant and intermediate contacts but, within the two-segment Euler–Bernoulli model, migrate toward a common near-tip position in the high-stiffness, hard-indentation limit (**Fig. 2b**). The following sections evaluate these predictions against experiment and finite-element simulation to determine which physical ingredients are required to capture the behavior of real AFM probes.

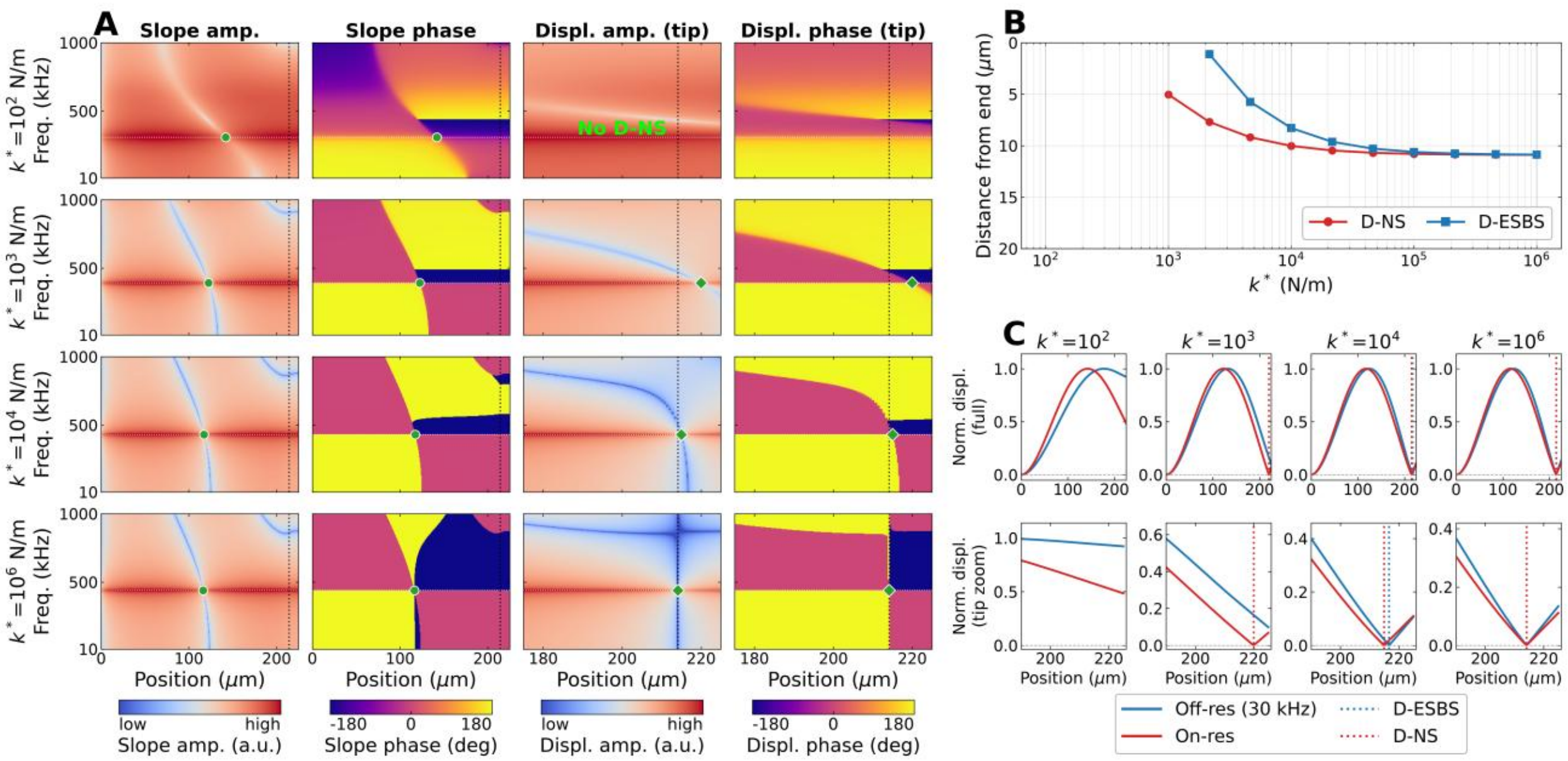

**Figure 2. Euler–Bernoulli model of null-spot and electrostatic-blind-spot dynamics for the Budget Sensors Multi75E-G probe, computed with the two-segment solver and an elastic-tip contact (isotropic $k_1 = k_2 = k^*$ acting through the tip cone, $k_{cone}$ = 2 kN/m).** (A) Frequency–position spectral maps at four contact stiffnesses ($k^*$ = $10^2$, $10^3$, $10^4$, and $10^6$ N/m): slope amplitude and phase over the full lever, and displacement amplitude and phase zoomed to the near-tip region. Horizontal dotted lines mark the first contact-resonance frequency, vertical dotted lines the physical tip position, and green markers the slope null spot (● S-NS) and displacement null spot (◆ D-NS); phase spectra are shown in addition to amplitude. (B) Normalized displacement profiles at the same stiffnesses

under on-resonance (red) and off-resonance 30 kHz (blue) excitation, with the D-NS and D-ESBS positions marked by dotted lines. (C) Migration of the D-NS and D-ESBS with contact stiffness (distance from the cantilever end); in this one-dimensional model the two conditions converge toward a common near-tip location at high stiffness, whereas the finite-element and experimental results (Figs. 3–5) retain a residual separation (Supplementary Note 1).

## Experimental Measurement of Null-Spot and Electrostatic Blind-Spot Dynamics

We now turn to experimental verification under realistic measurement conditions. Both operating conditions can be identified experimentally by translating the detection position along the cantilever: the NS is determined dynamically from the location where the resonant response vanishes (the intersection of the resonance and antiresonance branches in spatially resolved frequency sweeps), whereas the ESBS is determined quasistatically from the detection position at which the apparent piezoresponse contrast between oppositely polarized domains in periodically poled lithium niobate (PPLN) crosses zero. This zero-contrast criterion corresponds to the equal-amplitude method; because it balances the piezoelectric response against all bias-dependent contributions simultaneously, it is intrinsically sensitive to lateral and vector electromechanical signals in addition to the body-electrostatic force. An alternative determination based on minimum bias sensitivity (DC-ramp method) is illustrated in **Fig. S3**. In what follows, we focus on displacement detection, where the D-NS and D-ESBS both lie near the cantilever tip and their separation is most critical for quantitative interferometric PFM[10]. Because these operating conditions arise from different physical mechanisms, they must be identified independently under identical tip–sample boundary conditions.

To reproduce conventional IDS-PFM operations while systematically mapping these operating conditions, we developed the automated workflow shown in **Fig. 3a**. At each detection position along the cantilever and each applied load, two measurements are acquired sequentially: an electrically driven frequency sweep through the first contact resonance and a quasistatic PFM amplitude image across oppositely polarized PPLN domains. The resulting datasets enable simultaneous extraction of (i) the D-NS from the convergence of the resonance and antiresonance branches in the spatially resolved spectrogram and (ii) the D-ESBS from the detection position at which the trace- and retrace-domain contrast crosses zero. We applied this workflow to two substantially different interferometric probe designs—an Adama AD-2.8-AS diamond-tip probe (**Fig. 3**) and a platinum-coated BudgetSensors Multi75E-G probe (**Fig. 4**)—to determine how the D-NS, D-ESBS, and their separation evolve with loading and probe geometry.

**Figure 3b** shows representative spatially resolved resonance spectrograms for the Adama probe at three applied loads (250, 750, and 1250 nN), together with the corresponding domain-contrast bar charts below each spectrogram. The resonance branch —visible as the high-amplitude band whose frequency varies only weakly with detection position— stiffens systematically with increasing load, consistent with progression toward the hard-indentation regime. The antiresonance branch, by contrast, migrates strongly with detection position and intersects the resonance branch at the D-NS (red dotted line). The bar charts make the D-ESBS directly visible as the detection position at which the contrast between the two PPLN domain orientations crosses zero (black dashed line). For every load, the D-NS and D-ESBS occur at clearly different detection positions along the cantilever, in contrast to the high-stiffness convergence predicted by the two-segment beam model (**Fig. 2**).

The bar charts in **Fig. 3b** also quantify the local slope of the domain contrast with respect to detection position around the D-ESBS. For the Adama probe at the lowest load (250 nN) we measure a contrast gradient of approximately −0.12 pm/µm. Because this absolute value scales with the sample's piezoelectric response— it is more informative to normalize it by the local signal amplitude (≈5.5 pm), which yields a relative sensitivity of ≈2 %/µm. In other words, a one-micrometre error in the position of the detection laser introduces a ≈2 % error in the apparent piezoresponse. This gradient diminishes systematically as the load is increased: at 1250 nN it falls to ≈−0.09 pm/µm (≈1.6 %/µm), and an equivalent value of ≈0.02 pm/µm (≈0.4 %/µm) is observed for the BudgetSensors probe in Fig. 4b (which shows less sensitivity to applied load). This load dependence is consistent with the movement of the D-NS and D-ESBS away from the cantilever free end: in the hard-indentation regime the spatial separation between the two conditions becomes smaller and the local sensitivity of the apparent piezoresponse to detection-position error correspondingly decreases.

This sensitivity is compounded by the finite size of the detection laser spot, which is typically ≈4 µm in diameter. Even with perfect targeting the measurement averages the response over this footprint, across which the apparent signal varies by ≈8 % at the lowest load; and any residual offset of the spot from the true null translates directly into a systematic error at ≈2 %/µm. The finite spot size therefore sets a practical floor on how precisely the D-ESBS can be located and how completely the electrostatic contribution can be nulled. We note, however, that we observed variability between different probes (even from the same commercial batch), indicating strong sensitivity to minute changes in the tip–sample contact; we return to this point below.

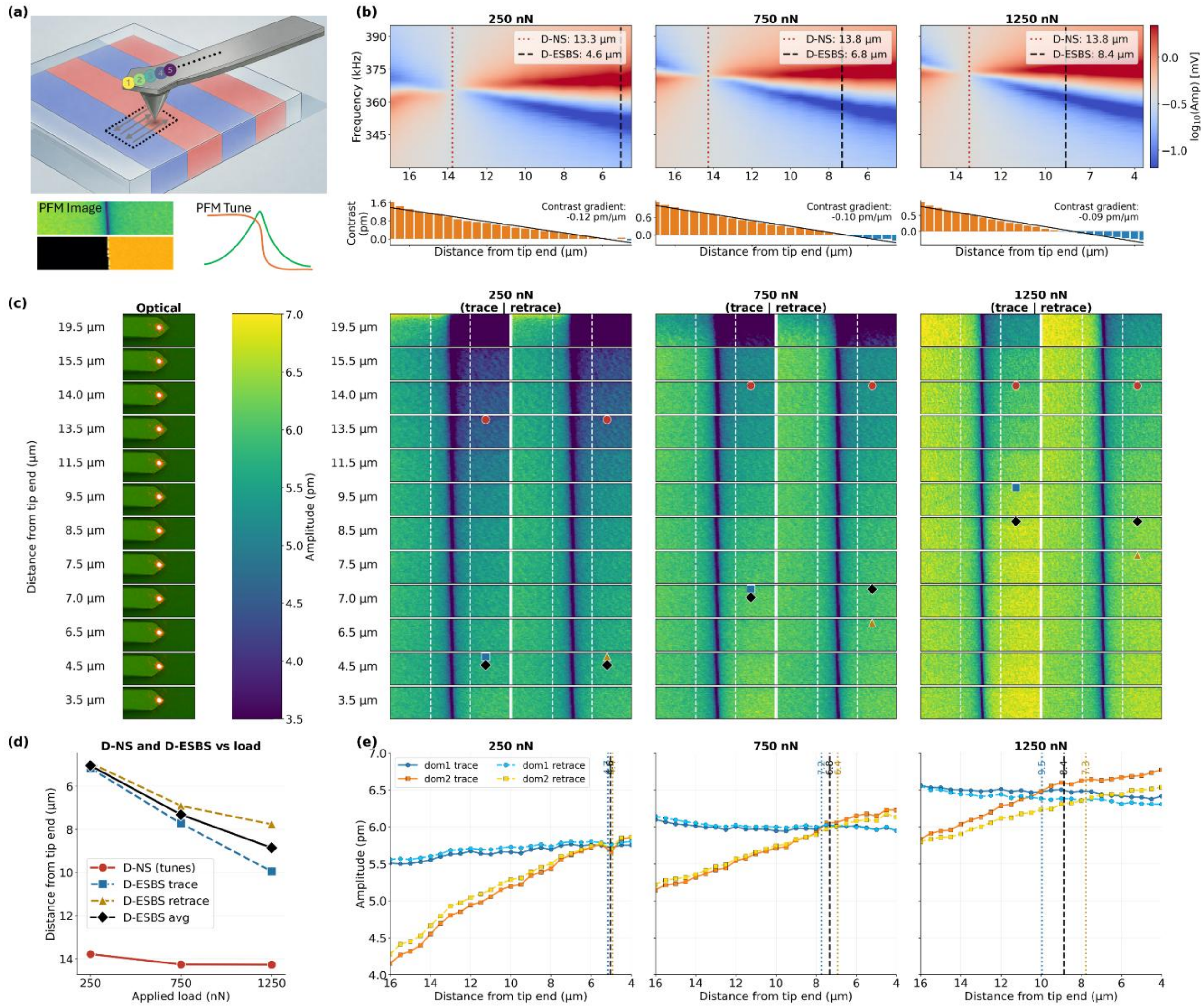

**Figure 3. Diamond-tip Adama probe — load-dependent mapping of D-NS and D-ESBS.** (a) Schematic of the experimental workflow: a PFM tip is scanned over PPLN domains under a controlled applied load while the interferometric detection spot is translated along the cantilever. At each cantilever position and each load (250, 750, and 1250 nN), a resonance frequency sweep and a quasistatic PFM amplitude image are acquired. (b) Spatially resolved resonance spectrograms (upper) and corresponding domain-contrast bar charts (lower) as a function of detection position along the cantilever, shown for each applied load. The D-NS (dotted red line) is identified from the crossing of the resonance and antiresonance frequencies; the D-ESBS (dashed black line) is extracted from the detection position at which domain contrast vanishes. (c) Representative optical micrographs (left) and PFM amplitude images (trace | retrace, right) at selected cantilever positions for each applied load. Symbols indicate detection positions corresponding to the D-NS (circle) and D-ESBS (square, triangle, and diamond for trace, retrace, and average, respectively). (d) Load dependence of the extracted D-NS and D-ESBS positions (trace, retrace, and average) for the Adama probe. (e) Domain-averaged PFM amplitude for the two PPLN domain orientations (trace and retrace) as a function of detection position at each applied load, from which the D-ESBS is identified as the crossing point between domain signals.

**Figure 3c** shows representative optical micrographs alongside the PFM amplitude images (trace and retrace) acquired at selected detection positions on the cantilever for each load. The symbol overlays (circle for the D-NS; square, triangle, and diamond for the D-ESBS extracted from trace, retrace, and the average, respectively) make explicit how the apparent PFM contrast between the two domains evolves as the detection laser is translated along the cantilever. Operating at the D-NS suppresses the resonant response but leaves a finite contrast between oppositely polarized domains, while operating at the D-ESBS produces vanishing domain contrast despite a non-zero resonant response—confirming that the two operating points are not equivalent in practice. Additionally, the D-ESBS position is strongly affected by lateral forces, hence the separate mapping of trace and retrace scan trajectories. This trace–retrace difference is consistent with friction-induced lateral forces at the tip–sample junction that reverse sign with scan direction; such friction contributions are difficult to capture in static beam models and constitute an additional accuracy limit on how closely the ideal hard-indentation regime can be approached in practice. In the present work we have always scanned at a 90° angle with respect to the cantilever axis, but we have observed that scanning at a 0° angle with respect to the cantilever axis can also have a strong effect on the D-ESBS determination.

**Figure 3d** summarizes the load dependence of the extracted D-NS and D-ESBS positions for the Adama probe. Both positions migrate further away from the cantilever end as the applied load increases, in qualitative agreement with the predictions of the Euler-Bernoulli and finite-element models. However, the two trajectories evolve at different rates and remain offset by several micrometers over the entire load range investigated. The D-ESBS shows a noticeably stronger load dependence than the D-NS, consistent with its sensitivity to the evolving spatial distribution of the electrostatic force and lateral and friction forces rather than to cantilever eigenmode dynamics alone.

**Figure 3e** shows the domain-averaged PFM amplitude for the two PPLN domain orientations, plotted as a function of detection position along the cantilever for each applied load; the D-ESBS is identified as the crossing point between the two curves. Beyond locating the D-ESBS, these curves reveal an additional effect: the magnitude of the measured electromechanical response—read directly from the amplitude of the two domain curves away from the D-ESBS—increases systematically with applied load, and the same trend is observed for the Platinum-coated probe in **Fig. 4e**. This load-dependent magnification of the apparent piezoresponse in the interferometric PFM has been quantitatively analyzed in recent work [20].

The same behavior is reproduced on a second, substantially different probe design: **Figure 4** shows the identical measurement protocol applied to the Platinum-coated BudgetSensors Multi75E-G probe over a slightly wider load range (250, 750, 1250, and 1750 nN). Quantitatively, the absolute D-NS and D-ESBS positions, their separation, and the magnitude of the domain-averaged amplitude all differ between the two probes, but the structure of the load dependence is preserved. The agreement between two probes from different manufacturers, with different stiffnesses and tip geometries, indicates that the spatial separation between the D-NS and D-ESBS is not an artifact of a particular cantilever design but a robust feature of contact-mode PFM.

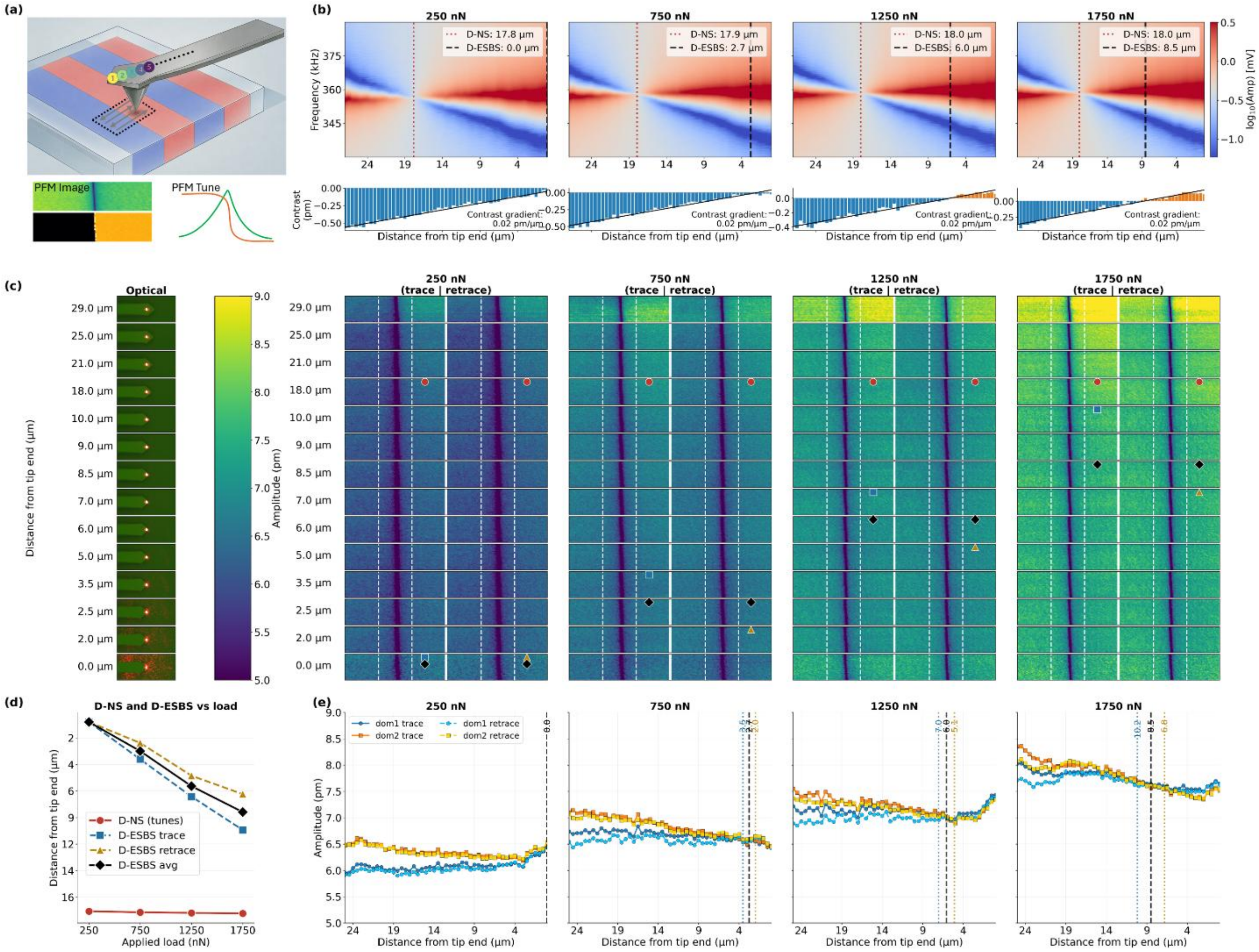

**Figure 4. Platinum-coated Budget Sensor probe — load-dependent mapping of D-NS and D-ESBS.** (a) Experimental workflow, identical to Fig. 3a, at applied loads of 250, 750, 1250, and 1750 nN. (b) Spatially resolved resonance spectrograms (upper) and corresponding domain-contrast bar charts (lower) as a function of detection position along the cantilever, shown for each applied load. The D-NS (dotted red line) is identified from the crossing of the resonance and antiresonance frequencies; the D-ESBS (dashed black line) is extracted from the detection position at which domain contrast vanishes. (c) Representative optical micrographs (left) and PFM amplitude images (trace | retrace, right) at selected cantilever positions for each applied load. Symbols indicate detection positions corresponding to the D-NS (circle) and D-ESBS (square, triangle, and diamond for trace, retrace, and average, respectively). (d) Load dependence of the extracted D-NS and D-ESBS positions (trace, retrace, and average) for the Budget Sensor probe. (e) Domain-averaged PFM amplitude for the two PPLN domain orientations (trace and retrace) as a function of detection position at each applied load, from which the D-ESBS is identified as the crossing point between domain signals.

Across all of these measurements we observed substantial probe-to-probe variability. Although this variability affects the absolute values of the D-NS, D-ESBS, and their separation, it does not alter the central observation that the D-NS and D-ESBS are distinct and spatially separated. SEM characterization of multiple tips—including individual probes drawn from a single commercial batch as well as nominally identical probes from different batches of the same model—reveals appreciable variation in cantilever length, tip height, cone angle, and apex geometry, even among probes sharing the same part number. Probe use can introduce further changes through wear, contamination, and evolving tip–sample contact conditions. This variability translates directly into the measured spectrograms, shifting the absolute positions of the resonance branch, antiresonance branch, D-NS, and D-ESBS between nominally identical probes. Consequently, the D-NS and D-ESBS positions reported in **Figs. 3** and **4** are specific to the probes measured; in practice, their absolute positions, load dependence, and separation must be re-determined for each probe under the conditions of the intended measurement.

As emphasized by Proksch and co-workers, quantitative interpretation of PFM requires that the contact stiffness substantially exceeds the cantilever stiffness so that the mechanical boundary condition approaches the pinned limit and the electromechanical response is not convoluted with cantilever compliance.[19] They define the hard-indentation limit as the regime in which the measured response becomes independent of applied load and approaches the intrinsic value (≈12 pm/V for PPLN).[19] Here we use a complementary, more general criterion that does not require a calibrated ferroelectric sample: we track the contact-resonance frequency as a function of load (**Fig. S4**) and identify the onset of frequency saturation as a probe-independent signature of the hard-indentation regime. Similar contact-resonance-versus-stiffness behavior has been reported by Rosenberger et al.[26], who showed that resonance frequency shifts can be quantitatively related to contact stiffness through a validated mechanical model. In our measurements, frequency saturation typically required loads exceeding ~3 μN, suggesting that the true hard-indentation regime may be difficult to reach in mechanically compliant materials such as polymers, ultrathin films, or two-dimensional systems without inducing surface damage (**Fig. S5**). Consistently, in the idealized limit (tip at the free end, zero tilt, hard contact) the extended beam model recovers the classical near-tip displacement null central to quantitative IDS-PFM[19]; introducing realistic tilt, tip set-back, and finite contact stiffness splits this single near-tip null into the distinct, load-dependent D-NS and D-ESBS conditions resolved in **Figs. 3** and **4**.

Beyond the Adama and BudgetSensors probes shown in **Figs. 3** and **4**, we assessed the generality of the D-NS and D-ESBS behavior by repeating selected measurements on a stiffer NCLPt cantilever (**Fig. S6**). All probes exhibited the same qualitative trends—load-dependent migration of the null position followed by saturation in the hard-indentation regime—but the absolute null locations differed substantially. In particular, for the NCLPt probe the experimentally accessible D-NS remained at the beyond the tip of the cantilever over the measured load range demonstrating that the near-tip detection conditions are strongly probe dependent and that an accessible interior D-NS is not guaranteed by every probe geometry. This behavior of the NCLPt probe is consistent with EB modelling as well as finite-element simulations (**Fig. S7**), which predict that the D-NS lies beyond the physical tip location owing to the probe's geometry and three-dimensional mechanics. This finding reinforces the central message of this section: the D-NS and D-ESBS are dynamic, probe-specific, parameter-dependent conditions, and reliable quantitative IDS-PFM requires their independent verification under the actual experimental conditions of the measurement.

Taken together, the measurements presented in **Figs. 3** and **4**, supported by the supplementary data in **Figs. S4**-**S8**, establish four central results. First, the D-NS and D-ESBS are spatially distinct under all probed conditions. Second, both positions migrate toward the cantilever tip with increasing applied load but at different rates, so their separation persists across the experimentally relevant range. Third, the apparent piezoresponse near the D-ESBS depends systematically on the tip-sample stiffness/compliance, with the local domain-contrast gradient diminishing and the measured piezoresponse growing as the contact stiffens. Fourth, probe-to-probe geometric variability further broadens the range of D-NS and D-ESBS positions encountered in practice, so probe-specific characterization is essential. Under realistic conditions, the measured D-NS-to-D-ESBS separation of a few micrometers translates into apparent picoscale piezoresponse, comparable to the intrinsic electromechanical response of many technologically relevant materials. Minimizing the resonant response (i.e., operating at the D-NS) or simply positioning the detection laser at the cantilever tip therefore does not guarantee suppression of electrostatic artifacts; reliable quantitative IDS-PFM requires independent identification of the D-ESBS under the relevant measurement conditions, as discussed below.

## Finite-element modelling of Null Spots and Electrostatic Blind Spots:

The preceding Euler–Bernoulli analysis and experimental measurements establish that resonance NS are dynamic quantities that evolve with the tip–sample boundary condition and therefore do not, in general, coincide with ESBS. Nevertheless, systematic differences remain between the simplified beam model and experiment, particularly in the absolute positions of the D-NS and D-ESBS and in their tendency to converge toward the tip location at high contact stiffness. These discrepancies likely arise from simplifying assumptions inherent to the one-dimensional Euler–Bernoulli framework, which idealizes the AFM probe as a rigid tip attached to a slender beam with lumped tip mass and finite tip height. Probe compliance is increasingly recognized across multiple length scales in AFM. At the atomic scale, high-resolution non-contact AFM has established that functionalized probe apices undergo lateral relaxation under tip–sample forces, fundamentally governing image formation and intramolecular contrast.[27,28] At the nanometer scale, flexible high-aspect-ratio CD-AFM probes have been shown to bend by several nanometers during imaging, producing measurable dimensional bias that requires explicit correction.[29] More recently, Chen *et al.* demonstrated that even conventional high-resolution silicon AFM probes exhibit measurable elastic bending of the tip cone under realistic imaging forces, with direct consequences for image formation.[30] Despite experimental evidence that probe deformation influences AFM measurements from the atomic to the micrometer scale, essentially all analytical PFM models continue to assume a perfectly rigid probe with fixed geometry and contact conditions. We therefore investigate whether realistic probe compliance contributes to the discrepancies between beam theory, finite-element modelling, and experiment observed here.

To explore this, we performed finite-element simulations using SEM-derived cantilever geometries. In **Fig. 5**, we show the simulations for cantilever dynamics of a Platinum-coated BudgetSensors Multi75E-G probe (equivalent modelling for Adama AD-2.8-AS probes with a conductive single crystal diamond tip can be seen in **Fig. S8**). Unlike the idealized Euler–Bernoulli approximation, the finite-element model explicitly incorporates the true cantilever shape, finite tip dimensions, and realistic tip placement using SEM images to build the simulated geometry (**Fig. 5a**).

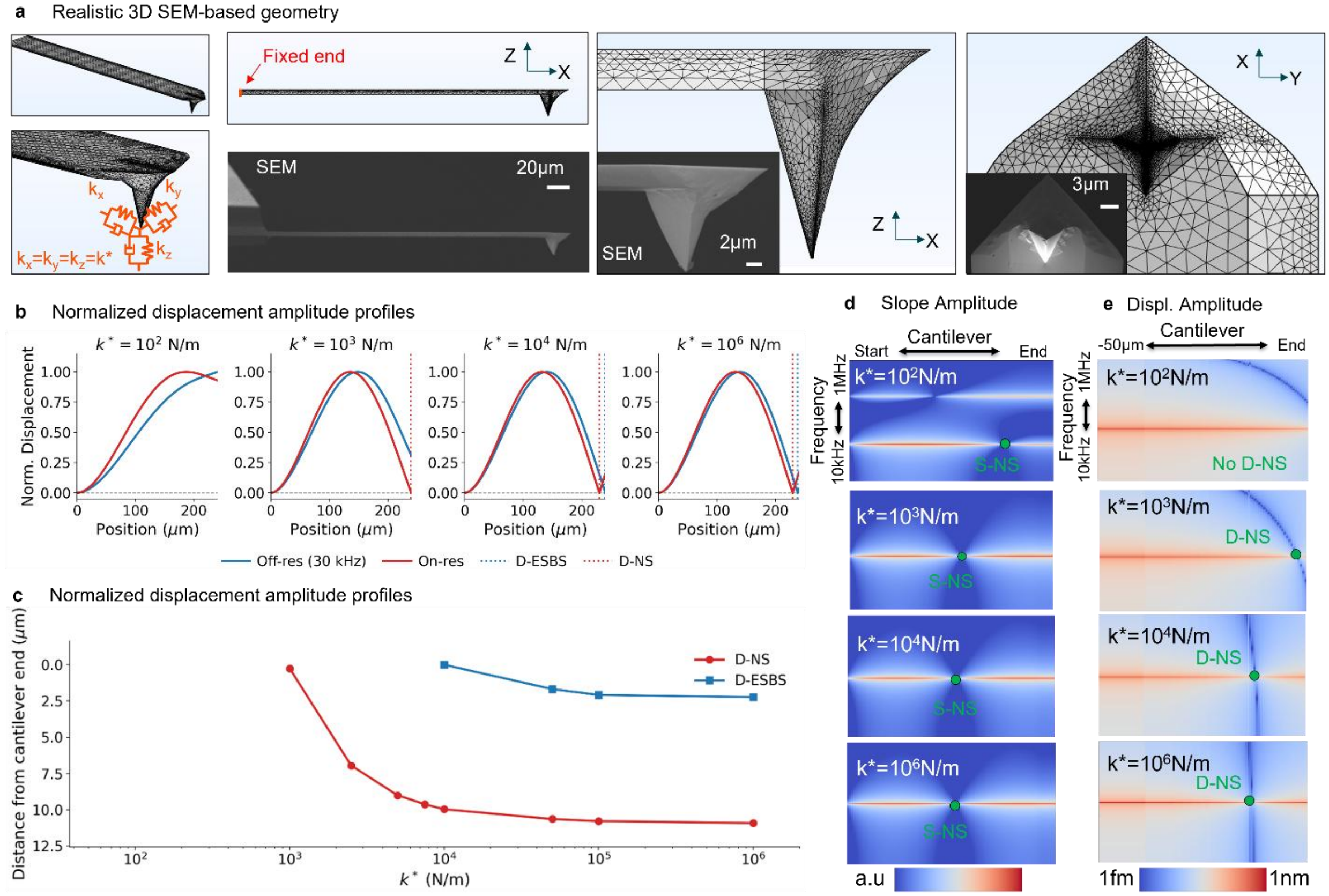

**Figure 5. Comparison of displacement null spots (D-NS) and displacement electrostatic blind spots (D-ESBS) predicted by realistic finite-element models.** (a) Finite-element simulations based on a realistic SEM-derived cantilever and tip geometry. SEM images and corresponding meshes are used to construct the model. (b) Normalized displacement amplitude profiles for different stiffnesses. (c) Distance from the cantilever end of D-NS and D-ESBS for different k*. (d) Simulated slope-amplitude frequency–position maps. (e) Displacement-amplitude frequency–position maps. Green markers indicate the locations of the slope null spot (S-NS) and displacement null spot (D-NS). Unlike the Euler–Bernoulli approximation, the realistic geometry produces persistent separation between the D-NS and D-ESBS, even at high contact stiffness, indicating that finite tip geometry and distributed electromechanical coupling prevent complete convergence of the two conditions in practical cantilever systems.

Representative displacement profiles and calculated D-NS and D-ESBS for increasing contact stiffness are shown in **Fig. 5b**. Consistent with the Euler–Bernoulli predictions of **Fig. 2**, both null positions migrate systematically with increasing contact stiffness and tend toward stable locations at high loads. However, unlike the Euler–Bernoulli model, the finite-element simulations predict that the D-NS and D-ESBS remain separated even in the nominal hard-indentation limit (**Fig. 5c**). This behavior closely mirrors the experimental observations of **Fig. 4**, where the two conditions remain distinct over the experimentally accessible load range and do not collapse onto the same detection position.

The improved agreement arises because the finite-element model captures physical effects absent from the one-dimensional Euler–Bernoulli approximation. Rather than representing the probe as a rigid tip attached to an ideal beam, the simulations incorporate the measured cantilever geometry, finite tip dimensions, realistic tip placement, distributed electrostatic loading, and three-dimensional elastic deformation derived directly from SEM images. As a result, the calculated mode shapes reproduce the experimentally observed near-tip behavior—including the pinned-like rebound near the contact (**Supplementary Note 3, Fig. S10a**)—that cannot be generated within the conventional beam framework.

Comparison with the beam model identifies the origin of these differences. In the Euler–Bernoulli formulation, the tip acts as a rigid kinematic constraint that couples beam displacement and slope at the contact. In a real probe, finite compliance of the tip structure relaxes this constraint, allowing relative motion between the cantilever and contacting apex and thereby modifying the local dynamic boundary condition. Analytical beam models also neglect the distributed electrostatic forces acting over the tip apex, cone, and cantilever overhang. Together, these limitations provide a consistent explanation for the experimentally observed differences in the absolute locations and convergence behavior of the D-NS and D-ESBS, consistent with growing evidence that probe deformation can influence AFM measurements across multiple length scales. [27-30] Matching the finite-element mode shapes requires realistic lateral contact stiffnesses (≈0.1–2 kN/m), yet under these conditions the beam model still predicts convergence of the D-NS and D-ESBS at high stiffness. Only by imposing an unrealistically rigid lateral constraint ($k_2 = k$)—which also suppresses the experimentally observed rebound (i.e. mode shapes no longer match)—can the one-dimensional model maintain a comparable separation. The persistent separation is therefore not simply a consequence of contact stiffness, but likely reflects three-dimensional probe mechanics and distributed electrostatic loading beyond the assumptions of the Euler–Bernoulli approximation (Supplementary Note 3).

## Conclusion

We demonstrate that the null spot (NS) and electrostatic blind spot (ESBS) are distinct PFM operating conditions arising from fundamentally different mechanisms. The NS is a dynamic property of the contact-resonance mode, whereas the ESBS results from cancellation of distributed electrostatic forces. Neither is therefore a fixed geometric location: both depend on contact stiffness, probe geometry, tip compliance, and the local electrostatic environment. The two-segment Euler–Bernoulli beam theory captures the dominant cantilever dynamics and qualitatively predicts the migration of these near-tip operating conditions. However, realistic finite-element simulations and experiments show that quantitative prediction of the D-NS and D-ESBS requires three-dimensional probe mechanics and electrostatic loading. Tip compliance and distributed deformation relax the rigid boundary condition assumed by beam theory, producing the persistent micrometer-scale separation observed experimentally.

These findings refine the conventional picture of interferometric PFM. Positioning the interferometer near the cantilever tip remains an effective strategy for reducing body-electrostatic artifacts, but the contact-resonance displacement null and electrostatic blind spot do not generally coincide. Micrometer-scale errors in detector placement can therefore introduce picometer-scale artifacts comparable to the intrinsic electromechanical response of fluorite ferroelectrics[31], ultrathin films and 2D ferroelectrics. For quantitative IDS-PFM, the electrostatic operating point should be identified independently of the mechanical null under the relevant measurement conditions. We have demonstrated such automated approaches herein.

The results also point toward opportunities for probe design. Our combined modelling and experiments identify probe geometry as a major contributor to the residual D-NS–D-ESBS separation. Eliminating the distal overhang and compensating the cantilever mounting angle so that the tip axis remains normal to the sample could reduce these geometric artifacts. Although the quantitative performance of such designs must be established through three-dimensional modelling and experiment, they provide a clear direction for future probe development (Fig. S11). More broadly, the principal challenge for next-generation quantitative PFM is not simply detecting ever-smaller displacements but faithfully understanding how the probe itself moves. As measurements approach the picometer and sub-picometer regime, the probe can no longer be treated as a passive detector—it must be considered as part of the measured response.

## Materials and Methods:

### Euler-Bernoulli model:

Cantilever dynamics were modeled using the analytical Euler–Bernoulli contact-resonance framework of Bradler *et al*[23]. The cantilever was treated as a tilted beam with finite tip height and elastic tip–sample contact, and the governing boundary-value problem was solved analytically to obtain frequency-dependent displacement and slope mode shapes. Because the tip of real probes is set back from the free end (SEM: $\Delta x \approx 10$–$20$ μm), the model was extended to a two-segment beam: the Bradler tip reaction enters as interior moment- and shear-jump conditions at the contact position, and the remaining overhang terminates in a genuine free end, $z''(L) = z'''(L) = 0$. The distributed electrostatic load acts on both segments, with the overhang carrying a reduced effective electrostatic width representing the tapered tip region. The implementation reduces exactly to the original tip-at-end model as $\Delta x \to 0$ and was verified against an independent finite-element discretization of the full beam (null positions agree to $\leq 3\times10^{-4}$ of the lever length). A simpler treatment — terminal boundary conditions applied at the contact with the deflection extrapolated across the overhang — violates the free-end conditions and produces spurious near-tip nodes; all results reported here use the two-segment solution (Supplementary Note 1, Figs. S14 and S15). A detailed comparison with prior treatments of the overhang — the eigenfrequency-only two-region analysis of classical contact-resonance microscopy and the tip-at-end forced-response models — is given in Supplementary Note 1. Simulations were performed over a range of contact stiffnesses representative of experimental conditions and were used to determine resonance frequencies, displacement null positions, and electrostatic blind-spot locations. The extended two-segment solver (CR2Beam / eb_cr_afm) and a notebook that regenerates all model figures are openly available at github.com/Liambcollins/EB-Solver-CResonance.

The tip–sample contact is represented by normal and lateral springs (with dashpots). Because a real silicon tip is not perfectly rigid, we optionally include a finite apex compliance — a lateral cone stiffness $k_{cone}$ — in series with the contact. This caps the rotational constraint ($\propto H^2 k_2$) that a stiff lateral contact would otherwise impose and releases the rigid-tip kinematic lock $z(x_c) = H \tan\theta\, z'(x_c)$ ($\theta = \varphi + \psi$), allowing the resonant (D-NS) and quasistatic (D-ESBS) near-tip nulls to occupy different positions. In the scalar limit this gives an effective lateral stiffness $k_{2,eff} = (1/k^* + 1/k_{cone})^{-1}$; the full treatment applies the same series combination through a 2×2 apex force balance (Supplementary Note 1). We fix $k_{cone} \approx 2$ kN/m by matching the on-resonance near-tip overhang rebound of the SEM-based finite-element model ($|z(L)|/\max \approx 0.17$; Fig. S10, Supplementary Note 3). The contact model used for each figure is listed in Table S1: the physical elastic-tip contact uses isotropic springs ($k_1 = k_2 = k^*$), and the frictionless ($k_2 = 0$) and rigid ($k_2 = k^*$) limits are used for the concept figure and for comparison.

Two excitation pathways are represented: a localized piezoelectric drive (and optional point force) at the tip apex, and a distributed body-electrostatic load $q(x) \propto V_{AC}\, \Delta V\, w_{eff}(x)/d(x)^2$ along the lever and overhang, where the local tip–sample gap $d(x) = H \cos\varphi + (x_c - x) \sin\varphi$ is set by the tilt and tip height. From the computed mode shapes, the displacement and slope null spots (D-NS, S-NS) are the on-resonance zeros of the piezo-driven response, and the displacement and slope electrostatic blind spots (D-ESBS, S-ESBS) are the quasistatic (30 kHz) zeros of the electrostatically driven response.

**Finite element model:**

We have used the Electrostatics and Solid Mechanics models of COMSOL Multiphysics 5.6 for realistic modeling which solves the Poisson equation using the scalar electric potential as the dependent variable and the fundamental equations of motion (conservation of momentum), which relate the deformation and movement of a material to applied forces, constraints and material properties. The realistic geometry of each tip is imported through an STL file into COMSOL. The STL file is built using custom-made software that inputs 3 SEM images of the tips used experimentally (see details of the meshing in Figure 5 and SI). The boundary conditions used were: the tip biased at a certain potential $V_{tip}$, the flat sample was grounded, the base of the cantilever is fixed (U=0) and the tip sample contact is simulated as 3 springs ($k_x$, $k_y$, $k_z$) that for simplicity were all set to the same value k*. We used a Si Young's modulus of 170 GPa for the tip material.

**Atomic Force Microscopy:**

We used a commercially available Vero AFM system (from Asylum Research, an Oxford Instruments company), equipped with a quadrature phase differential interferometer (QPDI). The movement of the interferometric laser (or the OBD laser) is automated through a custom Python script utilizing the AEcroscopy[32] platform. Tips used in different parts of the manuscript and SI were: Multi75E-G (BudgetSensors) with a free resonance frequency of 75 kHz and a spring constant of 3 N $m^{-1}$; and commercially available diamond probes from Adama model AD-2.8-AS with a free resonance frequency of 75 kHz and a spring constant of 2.8 N/m; and NCLPt with a free resonance of 190 kHz and a spring constant of 48 N/m.

**SEM Probe Characterization:**

After AFM measurements, probes were imaged using a ZEISS Merlin field-emission scanning electron microscope (FE-SEM) to determine cantilever and tip geometry. High-resolution SEM images were analyzed using a custom Python script that automatically extracted the cantilever length and tip position relative to the cantilever end. These experimentally measured geometric parameters were used as inputs for the Euler–Bernoulli contact-resonance model.

**Acknowledgements:**

This research was conducted at the Center for Nanophase Materials Sciences, which is a DOE Office of Science User Facility.

# Supplementary Information for: Dynamics of Null and Electrostatic Blind Spots for Quantitative PFM

Marti Checa[1*], Holland Swicegood[1], Ruben Millan-Solsona[1], Ralph Bulanadi[1], Stephen Jesse[1], Liam Collins[1*]

1 Center for Nanophase Materials Sciences, Oak Ridge National Laboratory, Oak Ridge, Tennessee 37831, USA.

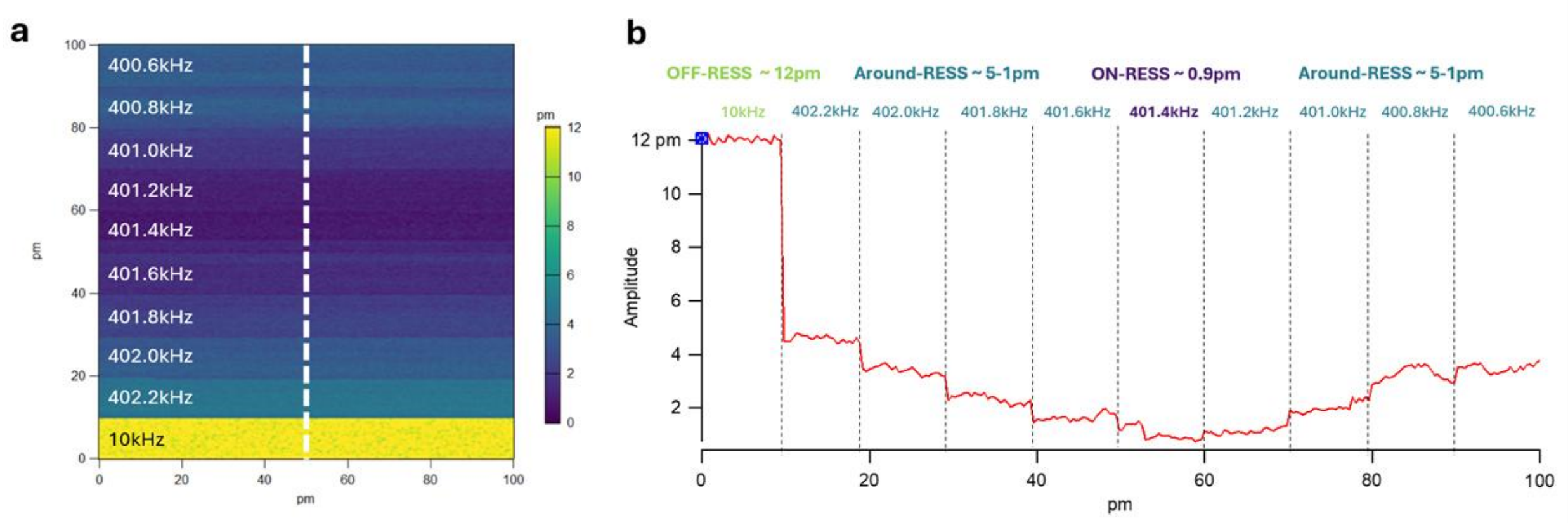

**Figure S1: Interferometric PFM amplitude at different frequencies measured on the D-NS (identified by nullifying the resonance peak).** (a) 100 pm scan (to avoid lateral forces acting on the tip), at varying frequencies from 10 kHz (quasistatic regime) to around the resonance (ruled by cantilever eigenmode). Data acquired at 3μN of applied load. (b) Profile along white dashed line. The PFM amplitude signal at the D-NS is reduced by ≈93 % relative to the same signal measured at 10 kHz and is practically at the detector noise floor.

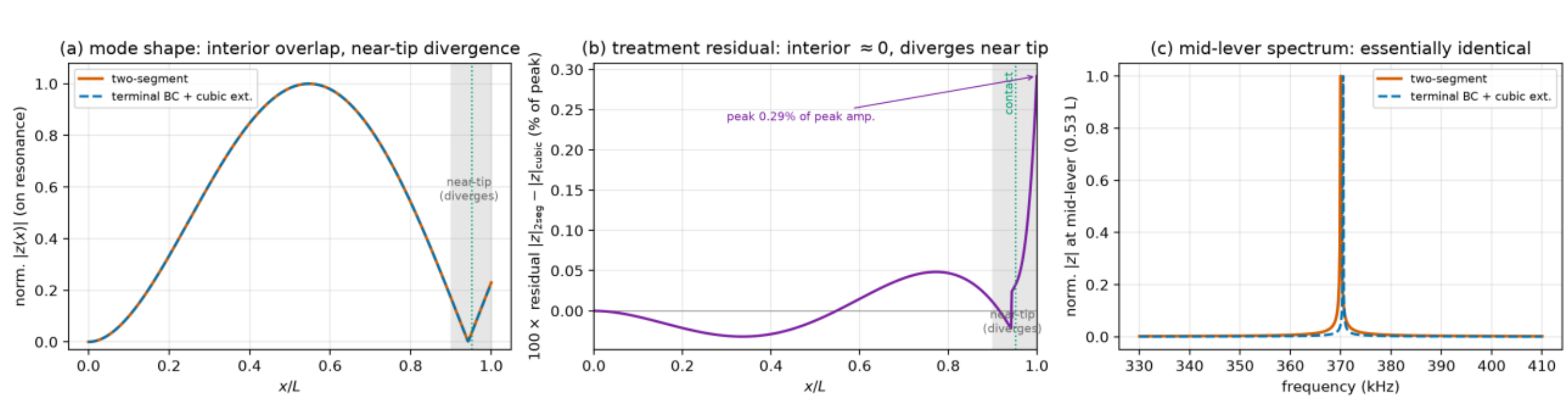

**Fig.S2. Two-segment overhang solution vs terminal-BC + cubic-extrapolation (Bradler-style) treatment.** Multi75E-G geometry, frictionless contact ($k_2 = 0$), $k^* = 10^4$ N/m. (a) On-resonance mode shape $|z(x)|$ normalized to its maximum. The two treatments overlap across the entire lever interior and part only over the shaded near-tip overhang ($x/L > 0.90$). (b) Signed residual between the two normalized mode shapes (×100, expressed as % of peak

amplitude): it stays below ~0.05% across the whole interior and rises to only 0.29% of peak over the last few percent near the tip. (c) Amplitude spectrum at mid-lever (0.53 L); the two treatments yield essentially identical contact resonance (f_CR differs by 446 Hz, 0.12%) and line shape. Together the panels show that the overhang treatment is largely immaterial for lever-body observables (Bradler's regime) but potentially important for the near-tip displacement nulls.

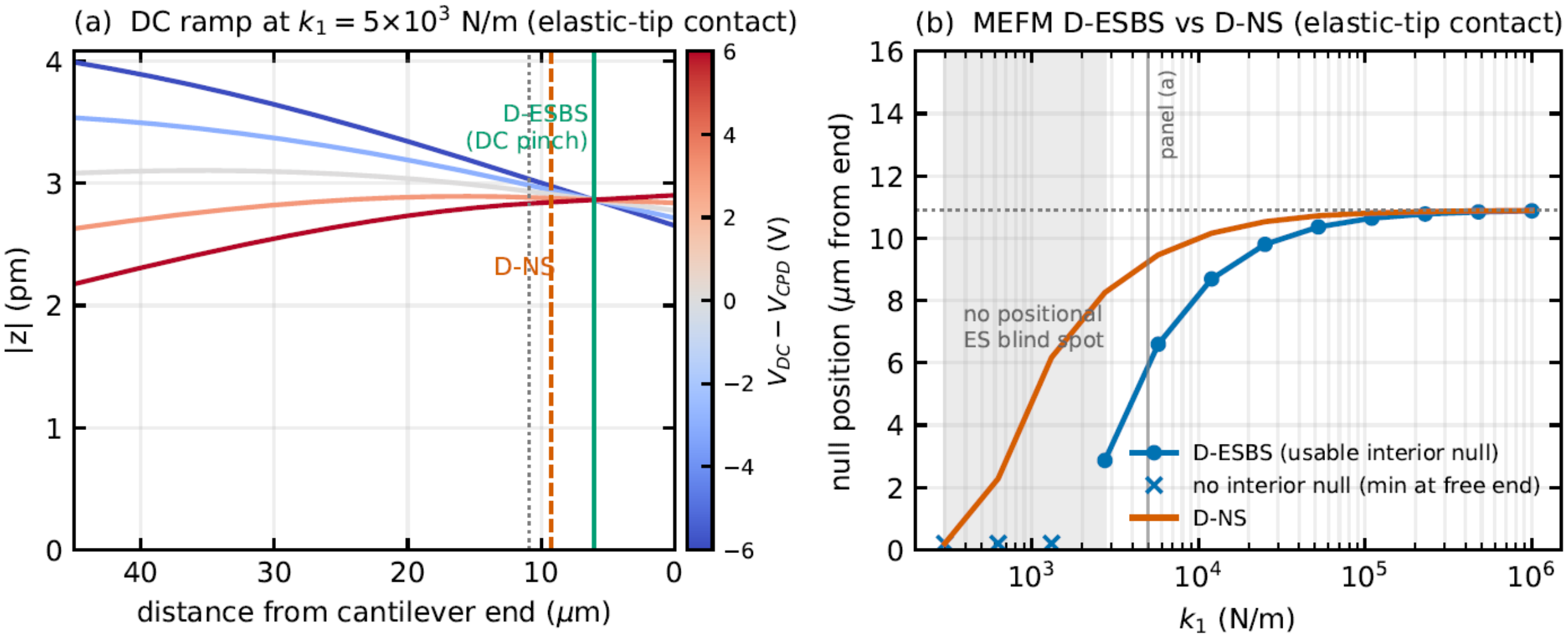

**Figure S3: Locating the D-ESBS by DC-bias ramp.** Multi75E-G geometry (L = 225 μm, overhang Δx = 10.9 μm, tip height H = 12 μm); lateral contact is the full elastic-tip model (isotropic springs $k_2 = k_1$ acting through a compliant tip cone, $k_{cone}$ = 2 kN/m). (a) Off-resonance tip displacement |z| along the lever at a fixed quasistatic drive (30 kHz) as the bias offset V_DC − V_CPD is ramped from −6 to +6 V (color scale) at $k_1 = 5\times10^3$ N/m. The ramped profiles cross at a single point — the DC pinch (green solid line) — which marks the D-ESBS, the position where the electrostatic contribution to displacement vanishes. The on-resonance displacement null (D-NS, orange dashed) sits ~3.2 μm away, closer to the contact point (gray dotted). At this moderate stiffness the two nulls are cleanly separated. (b) D-ESBS and D-NS positions versus lateral contact stiffness $k_1$. A usable interior D-ESBS exists only above $k_1 \approx 2.7$ kN/m; below that the electrostatic minimum runs off the free end (shaded region, × markers). As $k_1$ increases both nulls migrate to and converge onto the contact point, so the D-NS/D-ESBS separation is a low-stiffness feature of the 1-D model. The vertical gray line marks the stiffness used in panel (a).

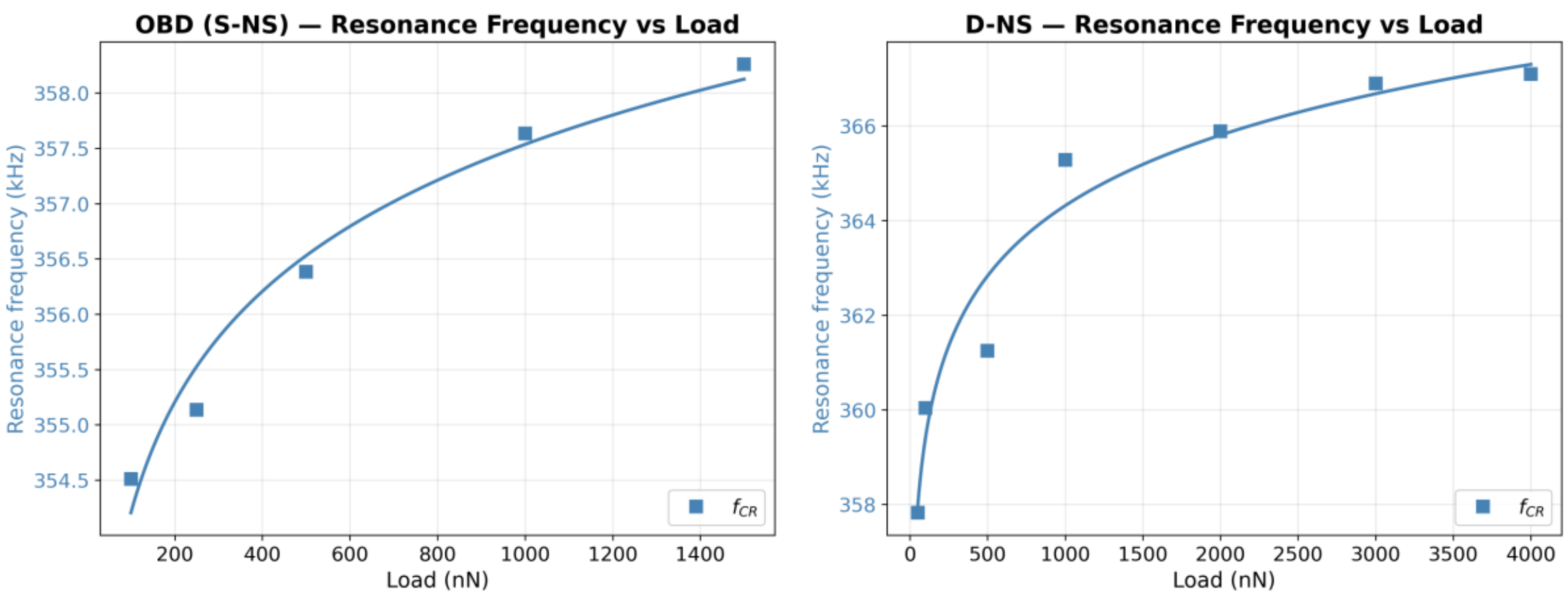

**Figure S4: Experimental contact-resonance frequency versus applied load.** Resonance frequency extracted from frequency sweeps acquired on a single PPLN domain at the indicated loads, using (left) optical beam deflection at the slope null spot (S-NS) and (right) interferometric displacement sensing at the displacement null spot (D-NS). Squares are measured values and solid lines are guides to the eye. Both curves exhibit the characteristic monotonic stiffening with applied load and approach saturation at high loads, consistent with progression toward the hard-indentation regime. This figure complements the discussion of Section "Experimental Measurement of Null-Spot and Electrostatic Blind-Spot Dynamics."

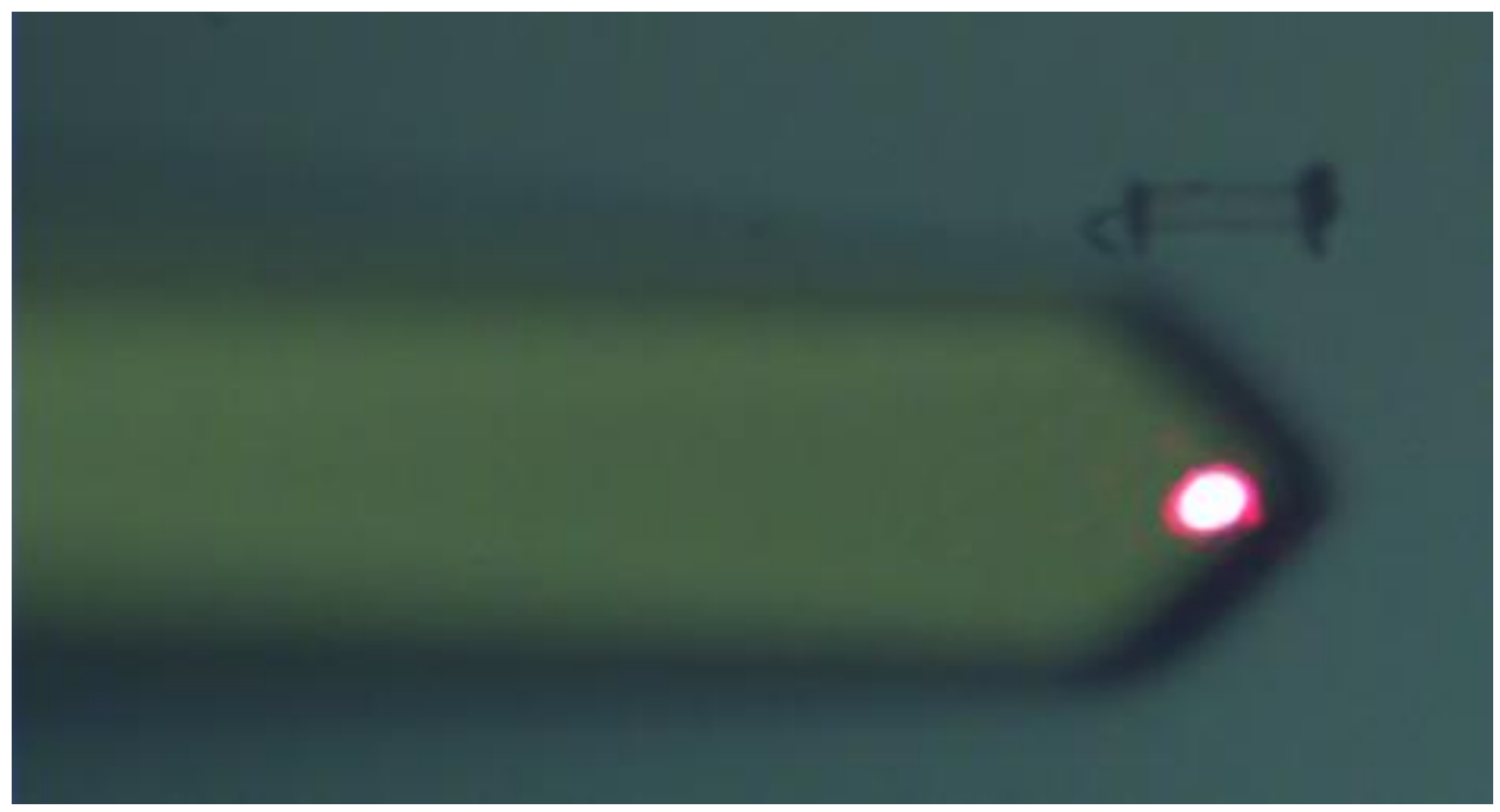

**Figure S5: Optical image after a 3μN scan, where damage to the PPLN sample can be observed.**

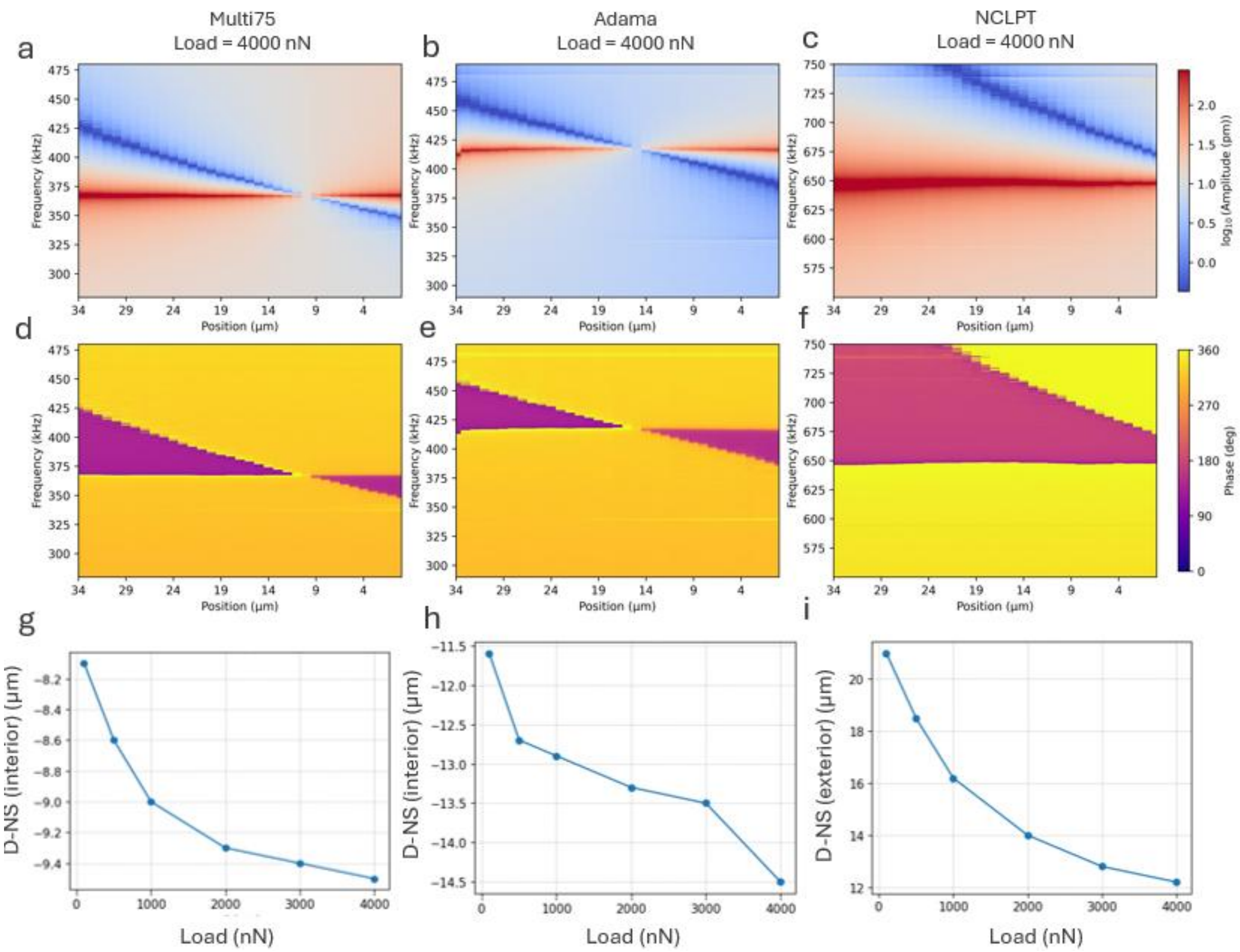

**Figure S6: Experimental determination and evolution of null-spot positions with tip–sample boundary conditions for three representative probes.** (a–c) Spatially resolved frequency–position maps of the cantilever transfer function measured using optical beam deflection (OBD) on PPLN at a load of 4000 nN for Multi75E-G, Adama, and NCLPt cantilevers, respectively. For the Multi75E-G and Adama probes, the contact-resonance and antiresonance branches evolve with detection position and intersect within the cantilever, defining an interior slope null spot where the resonant response is minimized. In contrast, the NCLPt probe does not exhibit a null spot within the physical cantilever length; extrapolation of the resonance–antiresonance crossing places the S-NS beyond the cantilever end. (d–f) Corresponding phase maps highlighting the resonance–antiresonance evolution and, where present, the crossing used to identify the null position. (g–i) Extracted S-NS position as a function of applied load for each cantilever; for NCLPt, the reported positions correspond to the extrapolated exterior null. Increasing load shifts the contact resonance to higher frequency and produces a systematic migration of the null position, demonstrating that the null spot is a dynamic property of the coupled cantilever–sample system rather than a fixed geometric location. The magnitude and direction of this shift depend on probe geometry and tip–sample boundary conditions, with all probes exhibiting progressive stabilization as the contact stiffness approaches the hard-indentation regime.

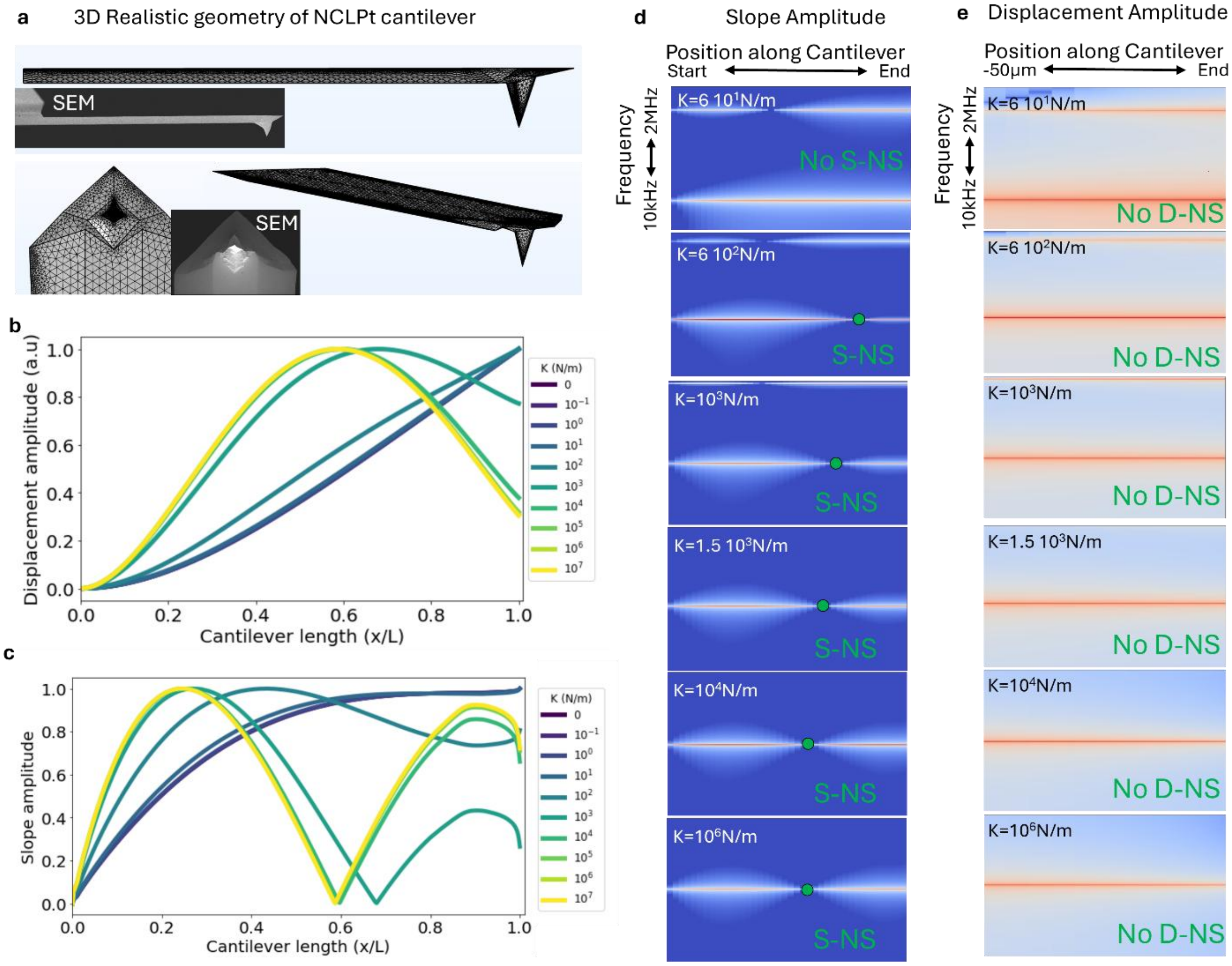

**Figure S7: COMSOL modelling of NCLPt tip geometry.** (a) Realistic 3D geometry of NCLPt tip extracted from SEM images. (b) Normalized displacement amplitude at different k*. (c) Normalized slope amplitude at different k*. (d) Frequency-position slope amplitude maps at different k*. (e) Frequency-position displacement amplitude maps at different k*.

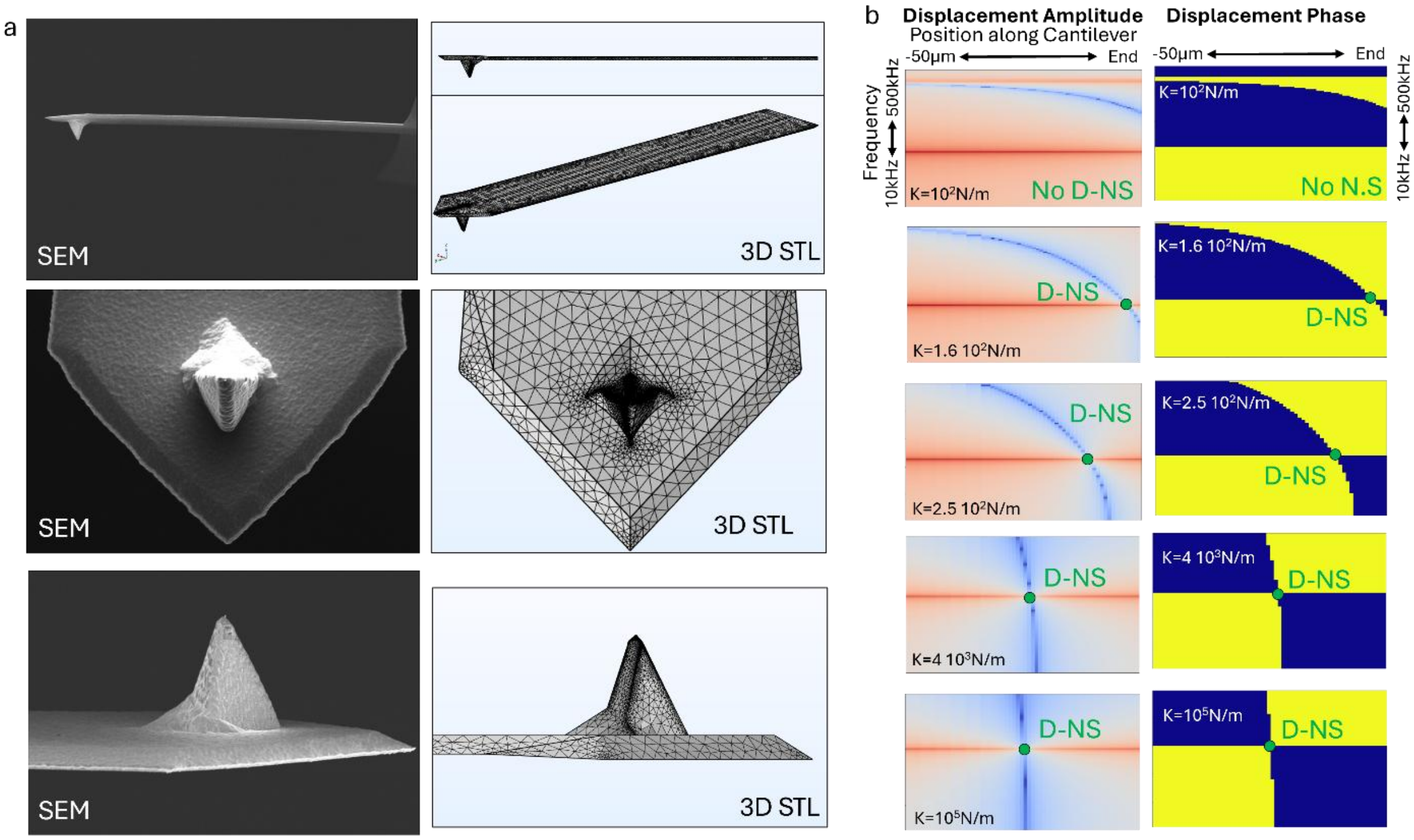

**Figure S8:** (a) Finite-element simulations based on a realistic SEM-derived cantilever and tip geometry of the Adama probe. SEM images and corresponding meshes are used to construct the model. (b) Displacement-amplitude (and phase) frequency–position maps. Green markers indicate the locations of the displacement null spot (D-NS).

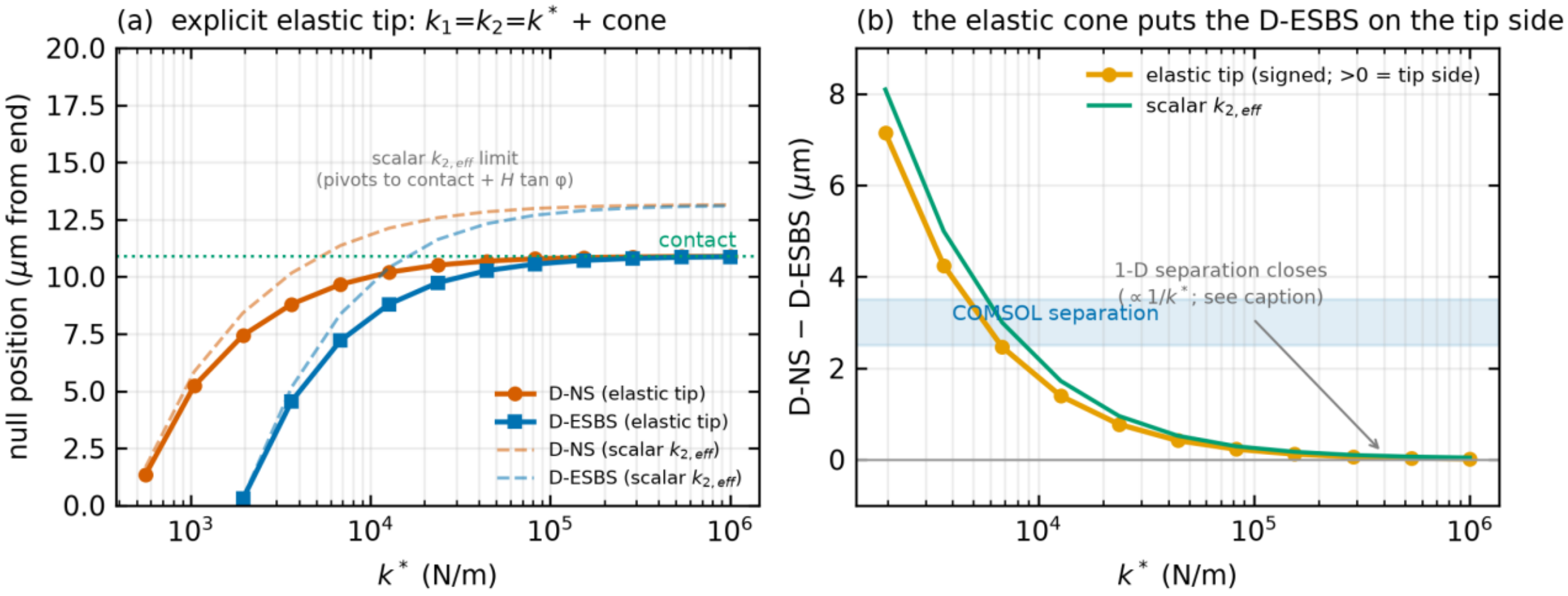

**Figure S9: Explicit elastic-tip contact. Raw isotropic contact springs ($k_1 = k_2 = k^*$, the finite-element condition) act through an elastic tip cone of lateral stiffness $k_{cone} \approx 2$ kN/m at the apex (the full 2×2 apex force balance; solid curves), compared throughout with the scalar compliance-limited spring $k_{2,eff} = \text{series}(k^*, k_{cone})$ (dashed).** (a) D-NS and D-ESBS positions versus $k^*$. The explicit elastic tip draws both nulls onto the contact point and — uniquely among the one-dimensional variants — places the D-ESBS on the tip side of the D-NS, matching the finite-element ordering with no overhang-width tuning. The scalar model instead pivots to a point $\approx H \cdot \tan \varphi$ beyond the contact, because it retains the rigid tip lever arm. (b) Signed separation D-NS − D-ESBS versus $k^*$ (positive = D-ESBS on the tip side). Both one-dimensional models pass through the finite-element (COMSOL) separation band at experimentally relevant stiffness, but the separation then closes as $\propto 1/k^*$ at high stiffness, as the axially rigid cone re-pins the beam displacement at the contact. The separation that persists in the finite-element model at hard contact therefore requires the three-dimensional electrostatic field loading the cone body — a drive channel absent in one dimension (Supplementary Note 3).

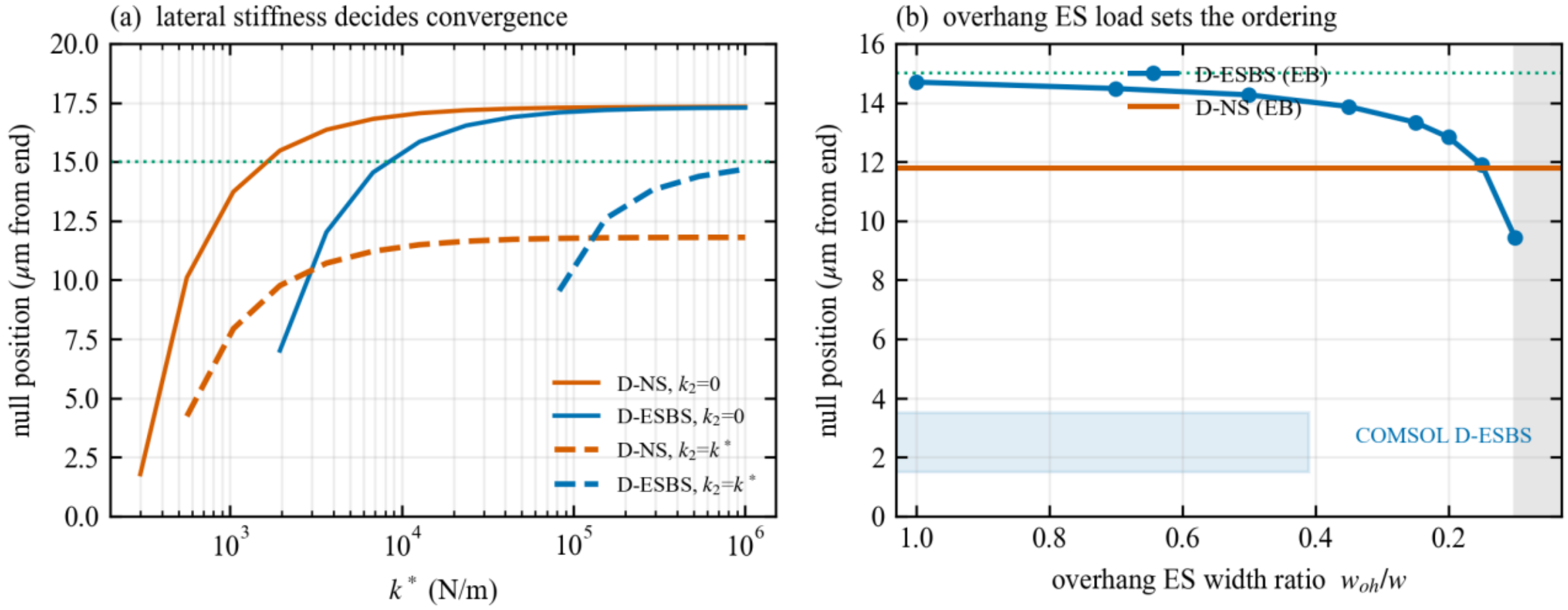

**Figure S10: Reconciliation with the finite-element results.** (a) On-resonance near-tip rebound |z(L)|/max versus lateral contact stiffness $k_2$ at $k^* = 10^5$ and $10^6$ N/m; the two curves coincide (the coupling is compliance-limited, not set by k*), and matching the COMSOL rebound band (≈0.17) fixes the effective cone stiffness at $k_{cone} \approx 2$ kN/m for this geometry (green dashed; $_{kcone} \ll k^*$ and geometry-sensitive — see Supplementary Note 1). (b) D-NS–D-ESBS separation versus contact stiffness: with the physically constrained lateral spring ($k_2 = (1/k^* + 1/k_{cone})^{-1}$) the two conditions converge at high stiffness, whereas only the unphysical rigid-tip limit ($k_2 = k^*$) reaches the COMSOL separation band — at the cost of the mode-shape agreement in (a). The persistent separation is thus not a lateral-stiffness effect but a three-dimensional electro-mechanical one: the elastic tip cone under three-dimensional electrostatic loading releases the rigid-tip z–z′ locking that pins all one-dimensional displacement nulls to a common point (Supplementary Note 1).

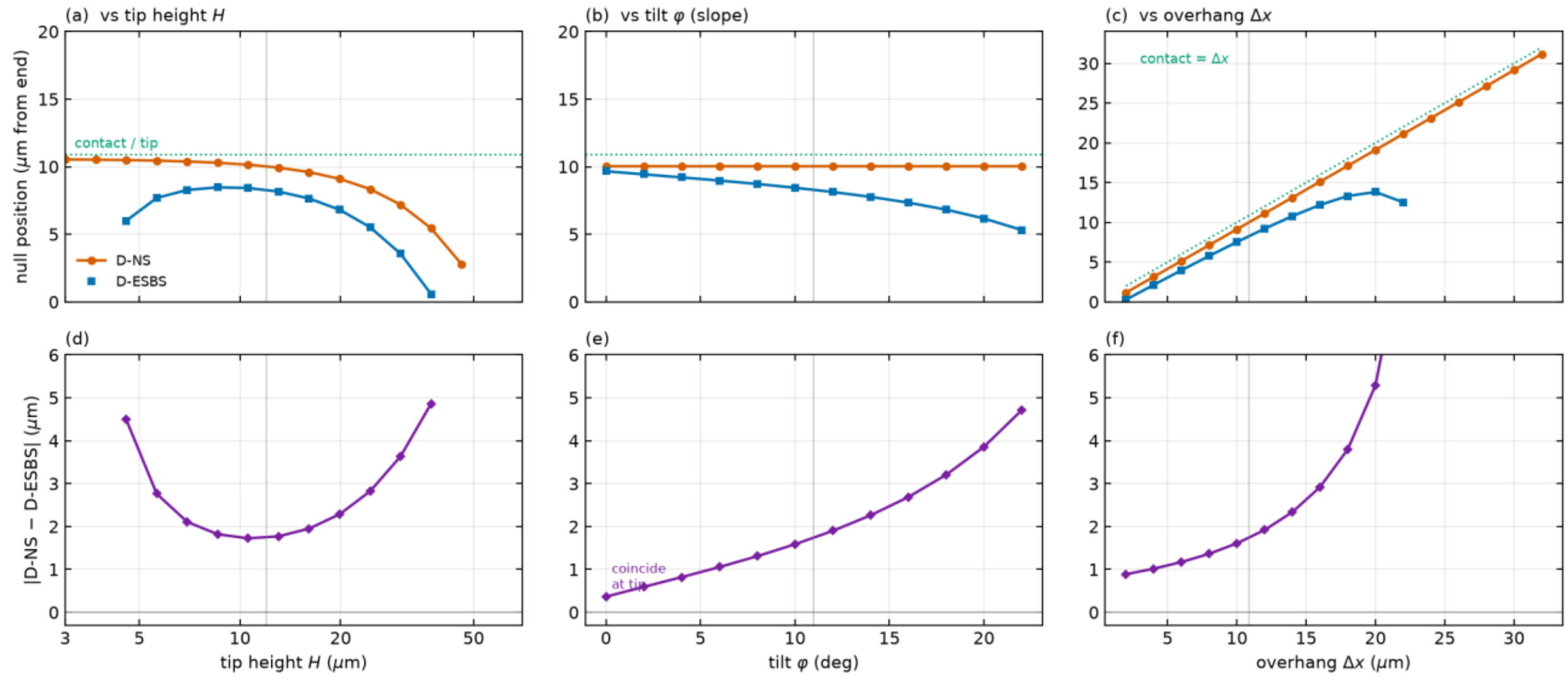

**Figure S11: Probe-geometry design study for a tip-coincident null. Top row: D-NS and D-ESBS positions (μm from the cantilever end); bottom row: their separation |D-NS − D-ESBS|.** Computed with the elastic-tip contact (isotropic $k_1 = k_2 = k^*$ acting through a compliant cone, $k_cone = 2$ kN/m) at $k^* = 10$ kN/m, varying one geometry parameter at a time about the Multi75E-G base (H = 12 μm, φ = 11°, Δx = 10.9 μm; gray line): (a,d) tip height H, (b,e) tilt φ, (c,f) overhang Δx. The green dotted line marks the contact/tip position (in (c) it moves with Δx). A single, well-defined operating point for quantitative PFM requires the two nulls to coincide at the tip (separation → 0). The separation is set almost entirely by the tilt: the D-NS is tilt-independent, while the D-ESBS shifts toward the tip as φ increases, so zero tilt brings them within ≈0.4 μm at the contact (b,e). An interior D-ESBS exists only for Δx ≳ 2 μm, and the separation then grows with overhang (c,f); the tip-height dependence has a shallow minimum near H ≈ 10–12 μm (d), close to the Multi75E-G value. For very tall tips (H ≳ 45 μm) or long overhangs (Δx ≳ 20 μm) the D-ESBS runs off the free end and no interior blind spot exists. The design direction for a tip-coincident null is therefore low tilt, a short but nonzero overhang, and moderate tip height — together with hard contact, at which all geometries draw the two conditions onto the contact point.

**Table S1. Cantilever, contact, and solver parameters used in the Euler–Bernoulli simulations.** All figures use the Budget Sensors Multi75E-G geometry; figure-specific contact stiffnesses are given in the table. Figure 1 uses the frictionless limit ($k_2 = 0$); Figure 2 uses an elastic-tip contact in which the isotropic springs act through the tip cone ($k_{cone} \approx 2$ kN/m for this geometry, of order the Si-cone lateral bending stiffness, fixed by the finite-element rebound; Fig. S10a). The compliance-limited scalar form $k_2 = (1/k^* + 1/k_{cone})^{-1}$ and the explicit elastic cone are described in Supplementary Note 3 (Figs. S9 and S10).

| Category | Symbol | Description | Multi75E-G (Figs. 1 & 2) |
|---|---|---|---|
| **Cantilever** | L | Total length | 225 μm |
| | w | Width | 28 μm |
| | H | Tip height | 12 μm |
| | k | Static spring constant | 2.8 N/m |
| | m | Total lever mass | ~52 ng* |
| | $m_t$ | Tip mass | 0 |
| | φ | Cantilever tilt angle | 11° |
| | Δx | Tip offset from free end | 10.9 μm (SEM) |
| **Contact** | $k_1$ | Vertical contact stiffness | 1500 N/m (Fig. 1); $10^2$–$10^6$ N/m (Fig. 2) |
| | $k_2$ | Lateral contact stiffness | 0 (Fig. 1, frictionless); k* through cone $k_{cone}$ ≈ 2 kN/m (Fig. 2) |
| | $c_1$ | Vertical contact damping | 20 μN·s/m |
| | $c_2$ | Lateral contact damping | 0 N·s/m |
| | ψ | Sample slope | 0° |
| **Solver** | Npts | Spatial grid points | 600 |
| | g | Air damping coefficient | 100 $s^{-1}$ |

*For Figure 2, the cantilever mass was back-calculated from the free resonance frequency (ffree = 75 kHz) using the Euler–Bernoulli beam relation (m = \beta_1^4 EI L/\omega_(free)^2), where (\beta_1 = 1.875/L).

†All Euler–Bernoulli simulations in the revised manuscript use the corrected two-segment solver of Supplementary Note 1. The overhang electrostatic width ratio w_oh/w (Supplementary Note 1) is 1.0 unless stated in a figure caption.

**Table S2. Excitation conditions used in the simulations.**

| Symbol | Description | Figure 1 Piezo | Figure 1 Body ES | Figure 1 Local ES | Figure 1 Combined | Figure 2 Piezo | Figure 2 ES | Figure 2 Combined |
|---|---|---|---|---|---|---|---|---|
| **$d_1$** | Substrate displacement amplitude | 3 pm | — | — | 3 pm | 3 pm | — | 3 pm |
| **VAC** | AC voltage amplitude | — | 3 V | — | 3 V | — | 1 V | 1 V |
| **ΔV** | DC contact potential | — | 1 V | — | 1 V | — | 1 V | 1 V |
| **$f_0$** | Local tip force amplitude | — | — | 100 pN | 100 pN | — | 1 pN | 1 pN |

**Supplementary Note 1: Extended two-segment Euler–Bernoulli model**

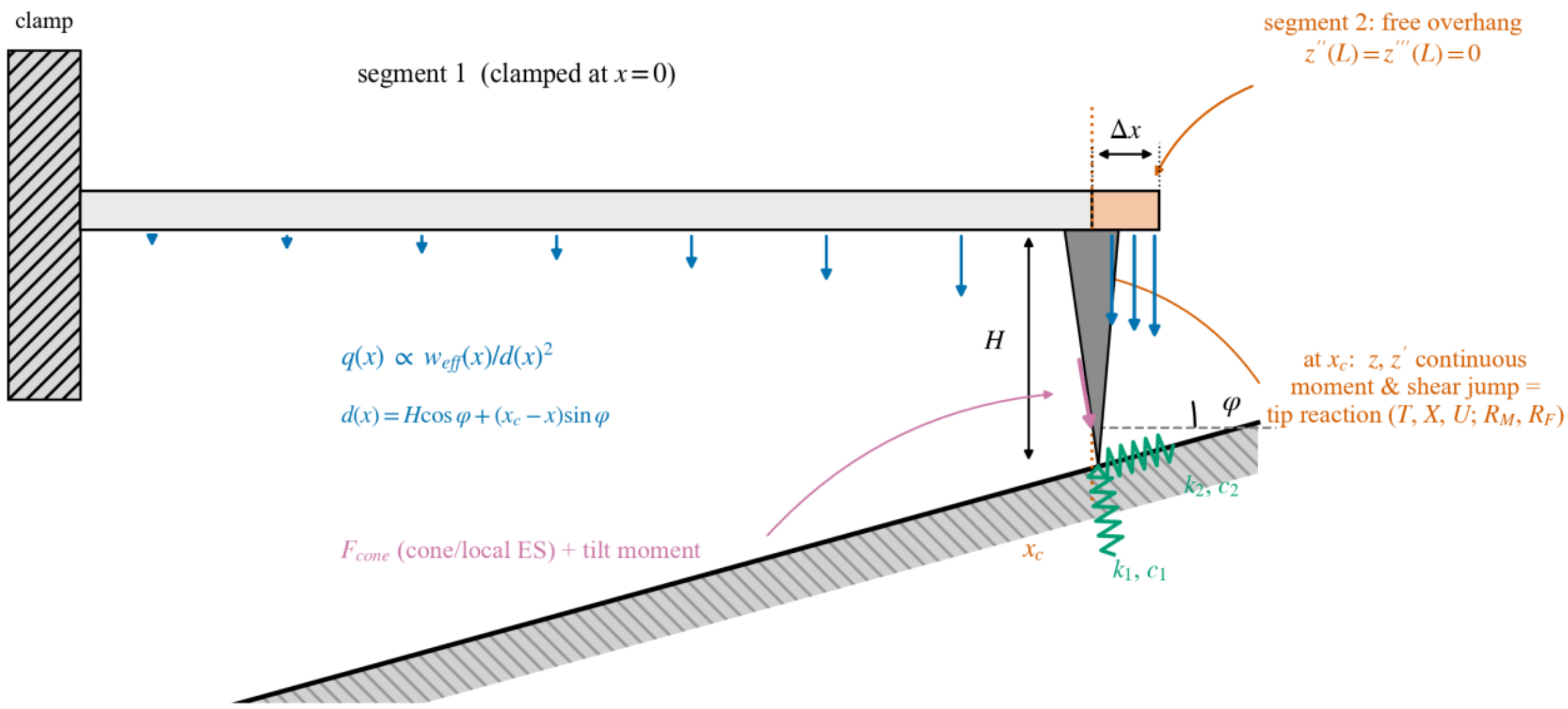

**Figure S12:** Schematic of the extended two-segment model: clamped tilted lever, tip of height H set back Δx from the free end, contact springs/dashpots at x_c, moment/shear jump conditions, free end at L, distributed electrostatic load on both segments, and cone-localized force at the tip.

This note builds the analytical model in three steps: (i) the original tip-at-end Euler–Bernoulli model of Bradler et al. (S1.1); (ii) a two-segment extension with an interior contact and a true free overhang (S1.2); and (iii) a compliant tip cone (S1.3). Section S1.4 summarizes the numerical solution and validation. Figure S12 shows the geometry.

**S1.1 Base model (Bradler et al. framework)**

The cantilever is a clamped, tilted Euler–Bernoulli beam with tilt angle φ, length L, flexural rigidity EI, mass per unit length m/L, and viscous damping coefficient g. In the frequency domain, the transverse deflection z(x) obeys

$$EI\, z''''(x) = \frac{m}{L}(\omega^2 - i\, g\, \omega)\, z(x) + q(x) \quad \text{(S1)}$$

or, equivalently, $z'''' = \alpha^4 z + q/EI$, where the complex flexural wavenumber is

$$\alpha^4 = \frac{m}{EI\, L}(\omega^2 - i\, g\, \omega) \quad \text{(S2)}$$

A rigid tip of height H and lumped mass mT contacts the sample through complex normal and lateral impedances

$$k_{1c} = k_1 + i\omega c_1, \quad k_{2c} = k_2 + i\omega c_2 \quad \text{(S3)}$$

where each impedance contains a spring k and dashpot c. The distributed electrostatic load and local lever–sample gap are

$$q(x) = -\cos\varphi \cdot \varepsilon_0 \varepsilon_r V_{AC}\, \Delta V\, w \,/\, [d(x)]^2, \quad d(x) = H\cos\varphi + (x_c - x)\sin\varphi \quad \text{(S4)}$$

Here w is the lever width and x_c is the tip position. In Bradler's original model the tip lies at the free end (x_c = L), so the contact enters through terminal boundary conditions written using three auxiliary functions T, X, and U:

$$\begin{aligned} z''(L) + T\, z'(L) + X\, z(L) &= R_M \\ z'''(L) - U\, z(L) - X\, z'(L) &= R_F \end{aligned} \quad \text{(S5)}$$

These functions convert the contact reaction at the tilted, finite-height tip into beam moment and shear. With θ = φ + ψ, where ψ is the contact-plane tilt,

$$T = \frac{H^2}{EI}(k_{1c}\sin^2\theta + k_{2c}\cos^2\theta) - \left(\frac{H}{4}\right)^2 \frac{m_T}{m}\, \alpha^4 L \quad \text{(S6)}$$

$$X = \frac{H}{EI}\sin\theta\cos\theta\, (k_{2c} - k_{1c}) \quad \text{(S7)}$$

$$U = \frac{1}{EI}(k_{1c}\cos^2\theta + k_{2c}\sin^2\theta) - \frac{m_T}{m}\, \alpha^4 L \quad \text{(S8)}$$

The excitation terms RM and RF are formed from the vertical force Fvert = k1c d1 + f0 and lateral force $F_{lon}$ = k2c d2, projected through the tilted tip:

$$\begin{aligned} R_M &= \frac{H}{EI}(\cos\theta\, F_{lon} - \sin\theta\, F_{vert}) \\ R_F &= -\frac{1}{EI}(\cos\theta\, F_{vert} + \sin\theta\, F_{lon}) \end{aligned} \quad \text{(S9)}$$

**S1.2 Interior contact and the overhang: two-segment extension**

Real probes place the tip about 10–20 μm before the cantilever end, leaving a free overhang of length L2 = Δx. Because Eqs. (S5) are terminal conditions, they cannot be applied directly at an interior contact. We therefore solve two beam segments: segment 1 from the clamp to x_c and segment 2 from x_c to the free end. The contact reaction appears as jumps in moment and shear at x_c, while the overhang satisfies the true free-end conditions:

$$\begin{gathered} z, z' \text{ continuous at } x_c \\ z''(x_c^-) - z''(x_c^+) + T\, z'(x_c) + X\, z(x_c) = R_M \\ z'''(x_c^-) - z'''(x_c^+) - U\, z(x_c) - X\, z'(x_c) = R_F \\ z''(L) = z'''(L) = 0 \end{gathered} \quad \text{(S10)}$$

As $\Delta x \to 0$, Eqs. (S10) reduce exactly to Bradler's terminal conditions because a zero-length overhang carries no moment or shear. The distributed load acts on both segments and is strongest near the free end, where the lever–sample gap is smallest.

Earlier implementations avoided a second beam segment by solving only to x_c, applying the terminal conditions there, and cubically extrapolating the contact-point state across the remaining overhang:

$$z(x_c + \xi) \approx z(x_c) + z'(x_c)\,\xi + \frac{1}{2}\, z''(x_c)\,\xi^2 + \frac{1}{6}\, z'''(x_c)\,\xi^3 \tag{S11}$$

This shortcut reproduces the lever-body response but fails near the tip. It carries the full tip-reaction moment and shear onto a segment that should carry only its own small inertia and distributed load, and it violates the free-end conditions $z''(L) = z'''(L) = 0$. The result is artificial displacement nodes on the same near-tip region where the physical D-NS and D-ESBS occur.

Figure S13 shows the consequence. The extrapolated solution creates a false second node and an artificial fold in the D-NS branch. Treating the overhang as a true second segment removes both artifacts and agrees with an independent finite-element solution (Fig. S14).

Despite this near-tip failure, the two approaches agree closely over the cantilever body: the resonant mode shape and mid-lever spectrum differ by less than 0.05% for $x/L < 0.9$ (Fig. S2). Cubic extrapolation is therefore adequate for lever-body observables, but not for near-tip null positions. All near-tip results use the two-segment model.

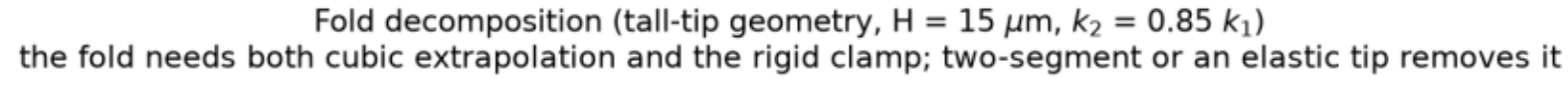

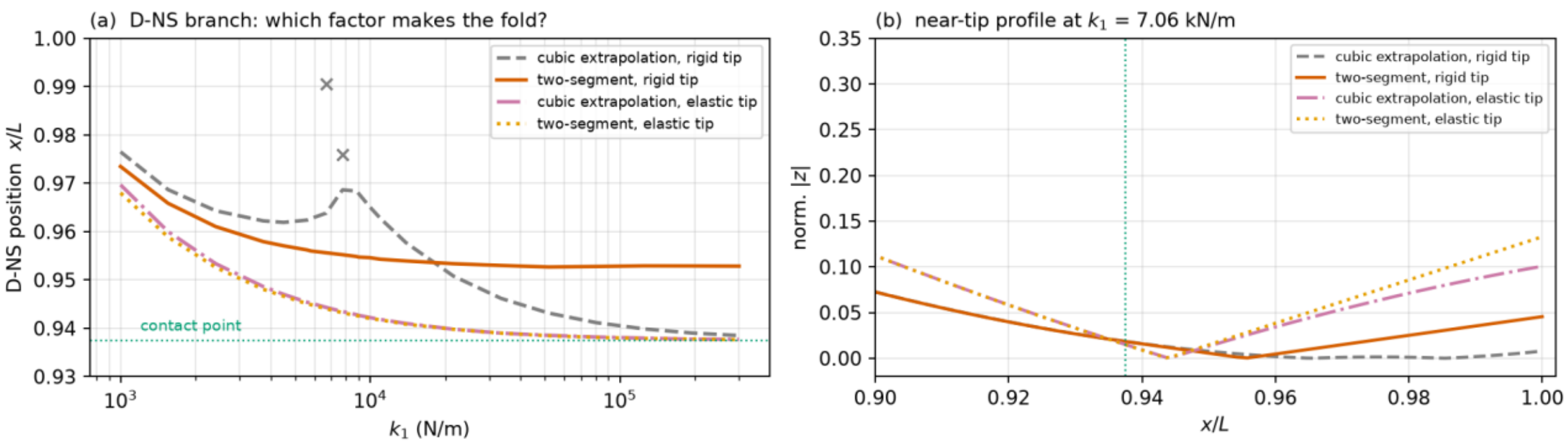

**Figure S13: Decomposition of the spurious D-NS fold.** Extracted D-NS position versus $k_1$ (tall-tip geometry, H = 15 µm) for the two-by-two combination of overhang treatment (cubic extrapolation, two-segment) and contact model (rigid tip, elastic tip cone at $k_{cone} = 2$ kN/m)tim, all at the same nominal springs $k_2 = 0.85k_1$. Crosses mark spurious secondary nodes. The fold and false second node appear only for the cubic extrapolation under the rigid rotational clamp (grey); capping the rotational constraint with the elastic cone removes the fold even with the extrapolation, and the two-segment treatment is smooth under either contact model.

### S1.3 Tip-cone compliance (elastic-tip contact)

The two-segment model still treats the tip as a rigid lever arm. The apex displacement u is therefore locked to the beam displacement and slope $q = (z, z')$ at the contact:

$$u = A\,q, \quad A = \begin{bmatrix} 0 & H \\ 1 & 0 \end{bmatrix}, \quad q = (z, z')^{\mathrm{T}} \tag{S12}$$

For a stiff contact, this gives the fixed relation

$$z(x_c) = H \tan\theta \cdot z'(x_c) \tag{S13}$$

which drives the resonant D-NS and quasistatic D-ESBS toward a common position. A real silicon tip is elastic. We represent its lateral and axial compliance by $C_{tip} = \mathrm{diag}(1/k_{cone,lat},\ 1/k_{cone,ax})$, replacing the rigid lock with an apex force balance:

$$u = A\,q + C_{\mathrm{tip}}\,f, \quad f = K_{\mathrm{c}}(d_{\mathrm{vec}} - u) + F_{\mathrm{ext}} \tag{S14}$$

where f is the total apex load, $K_c$ is the contact-stiffness matrix, $d_{vec}$ is the sample-side drive, and $F_{ext}$ is any direct apex force. Eliminating u gives

$$f = S\,(K_{\mathrm{c}} d_{\mathrm{vec}} + F_{\mathrm{ext}}) - S\,K_{\mathrm{c}}\,A\,q, \quad S = \left(I + K_{\mathrm{c}} C_{\mathrm{tip}}\right)^{-1} \tag{S15}$$

so the beam sees the series combination of the contact and cone stiffnesses, and the drive is filtered by the same operator S:

$$K_{\mathrm{eff}} = S\,K_{\mathrm{c}} = \left(K_{\mathrm{c}}^{-1} + C_{\mathrm{tip}}\right)^{-1} \tag{S16}$$

$$K_{\mathrm{c}} = k_{1\mathrm{c}}\,\hat{n}\hat{n}^{\mathrm{T}} + k_{2\mathrm{c}}\,\hat{t}\hat{t}^{\mathrm{T}} \tag{S17}$$

The auxiliary functions are then evaluated using Keff instead of the bare contact matrix. Setting $C_{tip} = 0$ ($k_{cone} \to \infty$) recovers Bradler's rigid-tip model. In most figures, the full 2×2 treatment is reduced to the scalar series stiffness

$$k_{2,\mathrm{eff}} = \left(\frac{1}{k*} + \frac{1}{k_{\mathrm{cone}}}\right)^{-1} \tag{S18}$$

which equals k* for a soft contact and saturates at kcone for a stiff contact. For the probes studied here, the lateral cone stiffness is of order 0.1–2 kN/m.

### S1.4 Numerical implementation and validation

Each beam segment is solved in closed form. Segment 1 uses the clamped basis; the overhang uses the full (cosh, sinh, cos, sin)(αξ) basis with ξ = x − x_c. Green's-function terms describe the distributed load, and a 6×6 linear system determines the modal coefficients. The solution gives the forced response, contact-resonance frequency, and full displacement and slope fields.

The implementation was checked in three independent ways. First, as Δx → 0, it converges to the Bradler terminal solution (relative field difference $\approx 10^{-4}$ at Δx = 1 nm). Second, an independent Hermite finite-element beam model reproduces the terminal solution to $\approx 10^{-5}$ and the two-segment null positions to within $3\times10^{-4}$ in x/L (Figs. S14 and S15). Third, an independent scipy solve_bvp calculation agrees with the closed-form fields to within 1%, with and without distributed electrostatic loading.

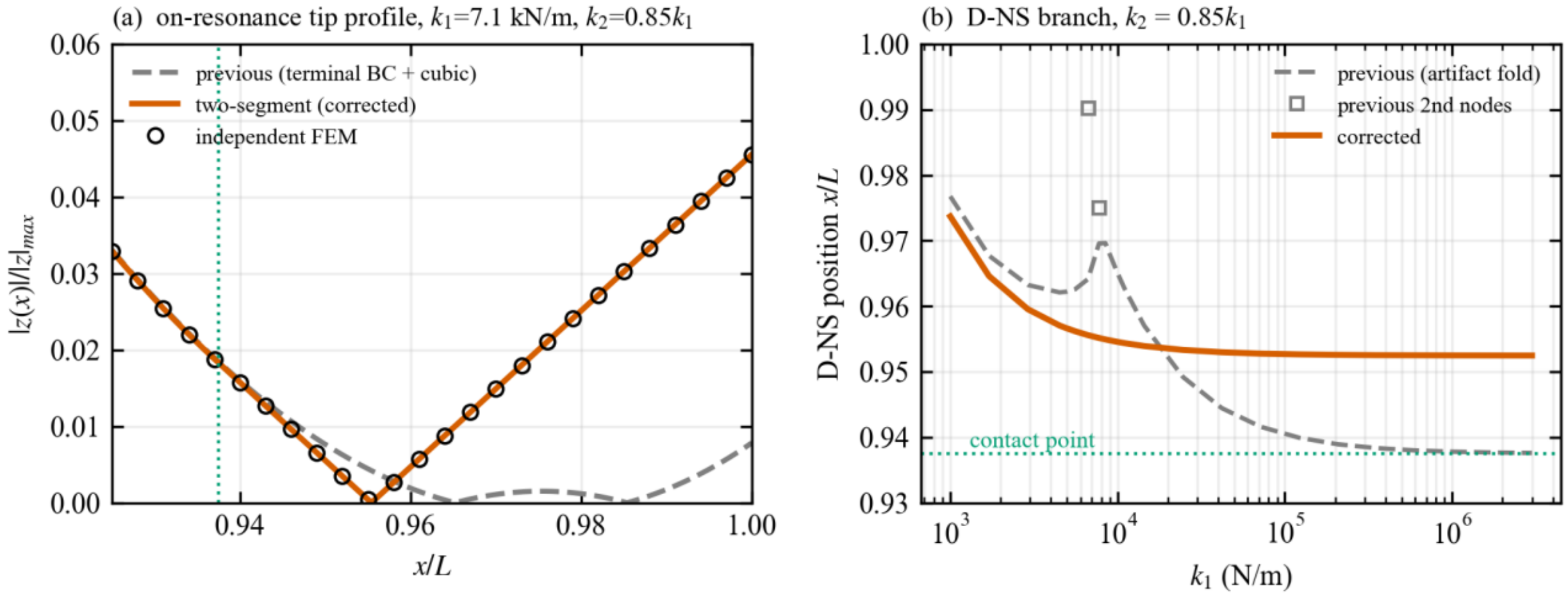

**Figure S14:** Previous (terminal-condition + extrapolation) versus corrected two-segment solution: (a) on-resonance near-tip displacement profile at $k_1 = 7.1$ kN/m, $k_2 = 0.85k_1$, with the independent finite-element check; (b) D-NS

versus $k_1$ showing the spurious fold of the previous treatment and the smooth corrected branch. The value $k_2 = 0.85$ $k_1$ here is chosen to expose the legacy solver's spurious fold at its strongest and is not the physical lateral stiffness used elsewhere (cf. Fig. S10).

COMSOL geometry: L = 240 μm, contact at $x_c/L$ = 0.955, green dotted). Both EB variants use the identical elastic-tip contact ($k_1$=$k_2$=$k^*$ through $k_{cone}$, rebound-matched for this geometry) — only the overhang treatm

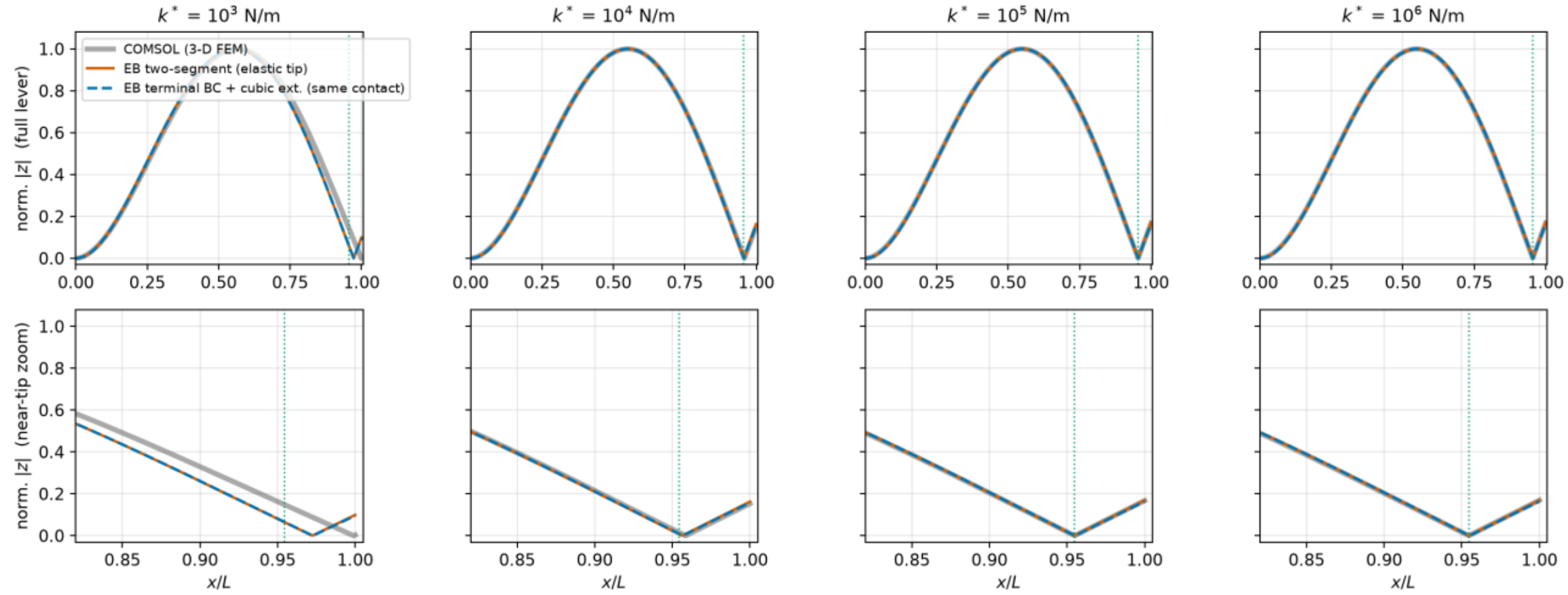

**Figure S15: Validation against the three-dimensional finite-element model.** On-resonance normalized displacement shape for the Multi75E-G probe on the finite-element geometry (L = 240 μm, contact at x/L = 0.955) at four contact stiffnesses ($k^* = 10^3$–$10^6$ N/m), full lever (top) and near-tip zoom (bottom): COMSOL (grey), the extended two-segment model with the elastic-tip contact (orange), and the terminal-boundary-condition-plus-cubic-extrapolation treatment using the identical contact (blue dashed). The two-segment model overlays the finite-element profile across the whole lever, including the near-tip node and the overhang rebound; with the physical (compliance-limited) contact the cubic extrapolation is nearly indistinguishable from it.

**Supplementary Note 2: Benchmarking four beam-level contact models — contact resonance, mode shapes, and near-tip null spots**

We compare four beam models to determine which assumptions affect contact resonance, mode shape, and near-tip null positions. M1 is the original tip-at-end model. M2 extends that solution across the overhang by cubic extrapolation. M3 solves the overhang as a true second beam segment. M4 adds finite tip-apex compliance to M3. Each model adds one physical ingredient, allowing us to separate effects of tip setback, free-end mechanics, and tip elasticity. The finite-element simulations and experiments in the main text provide the final tests of the beam-level predictions.

In M2, the beam is solved only on [0, x_c] and continued across the overhang by cubic extrapolation. In M3, the tip reaction acts at x_c through moment and shear jumps, and the overhang [x_c, L] ends freely with $z''(L) = z'''(L) = 0$. In M4, a lateral cone stiffness kcone = 2 kN/m acts in series with the contact springs and relaxes the rigid relation $z(x_c) = H \tan\theta\, z'(x_c)$. All models use the measured Multi75E-G geometry (L = 225 μm, Δx = 10.9 μm, φ = 11°), vertical piezo drive, and zero contact damping.

**Test protocol**

The normal contact stiffness k1 was swept from $10^2$ to $10^9$ N/m on a logarithmic grid. The tangential stiffness was set as a fixed fraction of k1: k2/k1 ∈ (0, 0.4, 0.85, 1) for M1–M3 and (0, 0.25, 0.5, 0.85, 1) for M4. At each point, the first contact resonance was located by mode continuation, and we recorded the full mode shape, the deepest near-tip displacement minimum, and the interior slope minimum.

Numerical checks included grid refinement, scaled and unscaled linear solves, a fine re-trace of the M3 high-stiffness branch, tracking of the first two modes to 2.2 MHz, and recovery of the clamped–pinned limit for $k2 = 0$. The null positions were stable to less than 0.003 µm, and no mode crossing or veering was found.

**Results**

Resonance (Fig. S16b). All four models agree at low contact stiffness and saturate above about $10^5$ N/m. M2 and M3 are spectrally identical to within 0.01%, showing that the overhang treatment has almost no effect on contact resonance. M4 reaches a lower plateau because the effective lateral stiffness saturates at kcone; once the contact is stiffer than the cone, the beam is insensitive to further increases in k2.

Displacement null (Fig. S16c). This observable separates the models. M1 cannot produce an interior near-tip null because the contact is placed at the beam end; at high stiffness the node lies on the boundary. M3 produces an interior null once k1 reaches about $3–6\times10^2$ N/m, followed by a broad plateau near 8.5 µm from the end. At very high stiffness (about $2–5\times10^7$ N/m), the rigid-tip constraint pushes this null smoothly through the free end. Fine-grid tracking confirms that this is a continuous branch termination, not a numerical jump.

M2 gives a different high-stiffness branch, including a false non-monotonic dip and a node created by the extrapolated tail; it also violates the free-end boundary condition. M4 removes the rigid-tip singular behavior: the null moves smoothly toward the contact for every tangential-stiffness ratio. Thus, within the beam models, the high-stiffness null position is controlled mainly by whether the tip is treated as rigid or compliant.

Slope null and mode shape (Fig. S16d). The slope null changes smoothly in all four models and provides little model contrast. The mode shapes are also nearly identical over most of the cantilever. The important differences are confined to the last ≈30 µm, where the models predict different null and rebound shapes.

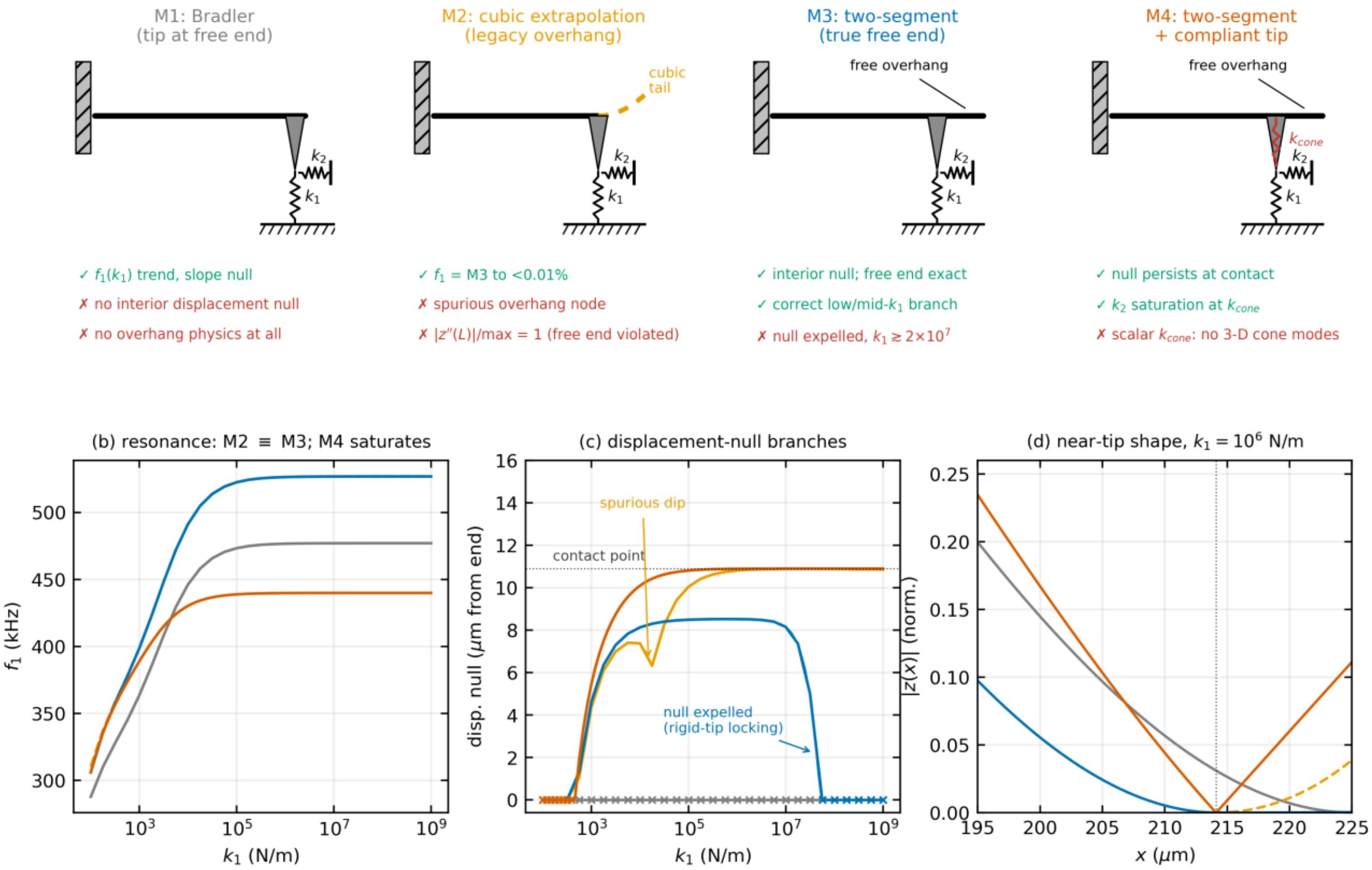

**Figure S16: Four-model benchmark.** (a) Model schematics with what each captures (green) and misses (red). (b) First contact resonance versus normal stiffness ($k_2/k_1 = 0.85$): M2 is spectrally identical to M3 (dashed); M4 saturates at 440 kHz. (c) Displacement-null branches: crosses mark boundary minima — a minimum at the beam end that does not count as an interior null. For M1 a genuine node does form at high $k_1$, but it coincides with the beam end because the contact sits at the end; only models with a tip setback (M3, M4) can host the null at an interior, observable position. The thin blue trace is the 46-point fine grid through the M3 expulsion; the dotted line marks the contact point at 10.9 µm from the end. (d) Near-tip mode shape at $k_1 = 10^6$ N/m; the dotted line is the contact point.

### What each model captures, what it misses, and why

| Model | Correctly captures | Misses / artifact | Suspected reason |
|---|---|---|---|
| M1 — Bradler, tip at free end (rigid tip) | $f_1(k_1)$ transition and saturation; slope-null trend; exact clamped–pinned limit for $k_2 = 0$ (336 kHz, +2.1% tilt excess) | No interior displacement null at any $k_1$ (the high-$k_1$ node coincides with the beam end); overstates end constraint; no overhang observables | The displacement null lives on the unloaded overhang beyond the contact; with zero tip setback that region does not exist, so the null cannot sit at an interior position |
| M2 — Bradler + cubic extrapolation (legacy) | $f_1$ identical to M3 to <0.01% at every point; interior mode shape up to the contact | Spurious overhang node; non-monotonic null dip (7.4→6.4 µm near $k_1 \approx 10^4$); free end violated: $\lvert z''(L)\rvert/\max = 1$ vs $\leq 10^{-16}$ for M3; agrees with M4 at high $k_1$ by accident | The cubic tail propagates the contact-reaction curvature and shear into a segment that must satisfy $z'' = z''' = 0$; the unloaded overhang carries only ~1% inertial moment, so the error is O(1), not perturbative |
| M3 — two-segment, true free end (rigid tip) | Interior null birth and low/mid-$k_1$ branch (~8.5 µm plateau); exact free-end conditions; correct overhang rebound topology | Null expelled through the free end for $k_1 \gtrsim 2\times10^7$ N/m (smooth on a fine grid — physical, not numerical); high-$k_1$ plateau likely too high (527 kHz) | Rigid-lever kinematic lock $z(x_c) = H\cdot\tan\theta\cdot z'(x_c)$: under a stiff tangential constraint $z(x_c)$ cannot vanish independently of the slope, so the null is pushed off the beam; the same lock |

| | | | overstates the rotational stiffening |
|---|---|---|---|
| M4 — two-segment + compliant tip ($k_{cone}$ = 2 kN/m) | Null persists at the contact point (10.9 µm) to $k_1 = 10^9$; smooth branches for all $k_2/k_1$; $k_2$ saturation at $k_{cone}$ (all ratios ≥ 0.25 collapse to 440 kHz) | Scalar, load-independent $k_{cone}$; no 3-D cone bending modes; by construction cannot reproduce the full D-NS/D-ESBS separation seen in FEM together with the observed rebound | A point spring at the apex lumps the distributed elasticity of the cone; the real cone bends and rotates, giving a drive- and frequency-dependent constraint that a single calibrated stiffness cannot represent |

Table S3: Scorecard of the four models against the sweep tests. Reasons in the last column are our interpretation of the failure mechanism.

**Interpretation**

Each model fails for a different reason. M1 has no overhang, so an interior displacement null has nowhere to form. M2 reproduces the lever interior but reconstructs the overhang from the wrong boundary data, creating artificial near-tip nodes. M3 treats the overhang correctly, but its rigid tip imposes an overly strong displacement–slope constraint at extreme stiffness. M4 relaxes this constraint with a single cone spring, giving smooth and persistent null branches, but it still cannot represent the distributed, drive-dependent deformation of a real three-dimensional tip. Thus, M1 is sufficient for many lever-body spectra, M3 captures the near-tip overhang topology, M4 captures the main effect of finite tip compliance, and full finite-element modelling is required for the coupled three-dimensional probe response.

**Practical consequences**

• Contact-resonance spectra alone cannot distinguish the overhang treatments; near-tip null positions can.

• A high-stiffness null at the contact or beyond the free end distinguishes compliant- and rigid-tip beam idealizations.

• Tangential contact stiffness is observable only below the cone-compliance limit.

• Dense stiffness sweeps are needed near null birth; coarse sampling can make a continuous branch appear to jump.

**Supplementary Note 3: Comparison with the three-dimensional finite-element model**

The beam models reproduce the main trend: both the D-NS and D-ESBS move with contact stiffness toward stable near-tip positions. They do not, however, reproduce the persistent separation seen in the SEM-based finite-element simulations and experiments at high stiffness. The finite-element model includes the measured cantilever and tip geometry, realistic tip placement, three-dimensional probe deformation, and the local electrostatic field. It also reproduces the pinned-like near-tip rebound observed experimentally (Fig. S10a).

Probe compliance is well established in AFM, from lateral relaxation of functionalized apices to bending of high-aspect-ratio and conventional silicon tips.[27-30] Most analytical PFM models nevertheless treat the tip as perfectly rigid, leaving its contribution to picometer-scale measurement errors largely unexplored.

In the Euler–Bernoulli model, a rigid tip imposes the lock $z(x_c) = H \tan\theta \, z'(x_c)$ (Eq. S13), which drives the D-NS and D-ESBS together in the hard-contact limit. Finite apex compliance partially releases this lock and allows the two conditions to occupy different near-tip positions.

We set the effective cone stiffness by matching the resonant overhang rebound to the COMSOL result. For the Multi75E-G geometry, $|z(L)|/\max \approx 0.17$ gives $k_{cone} \approx 2$ kN/m. With isotropic contact springs acting through this elastic cone, the one-dimensional model reproduces the finite-element ordering of the D-NS and D-ESBS and maintains a small separation over the experimental stiffness range (Fig. S10).

No physically reasonable one-dimensional contact reproduces both the finite-element mode shape and the full separation. When the model matches the observed rebound, the predicted separation is only 0.03–0.2 µm; a separation near 3 µm appears only for an almost rigid contact that suppresses the rebound (Fig. S17). The persistent separation is therefore not explained by lateral contact stiffness alone and instead reflects three-dimensional electro-mechanical coupling.

With the compliance-limited contact, the two-segment model closely matches the COMSOL resonant displacement shape, including the near-tip node and overhang rebound (Fig. S15). The remaining D-NS/D-ESBS separation arises from the three-dimensional electric field acting on the tip cone, a drive pathway absent from the one-dimensional model. The beam model is therefore sufficient for the mainly mechanical D-NS, whereas reliable prediction of the D-ESBS requires finite-element modelling. Under soft contact, a slope null may also exist when no displacement null is present, suggesting an advantage for slope-based detection in compliant or weakly contacting materials.

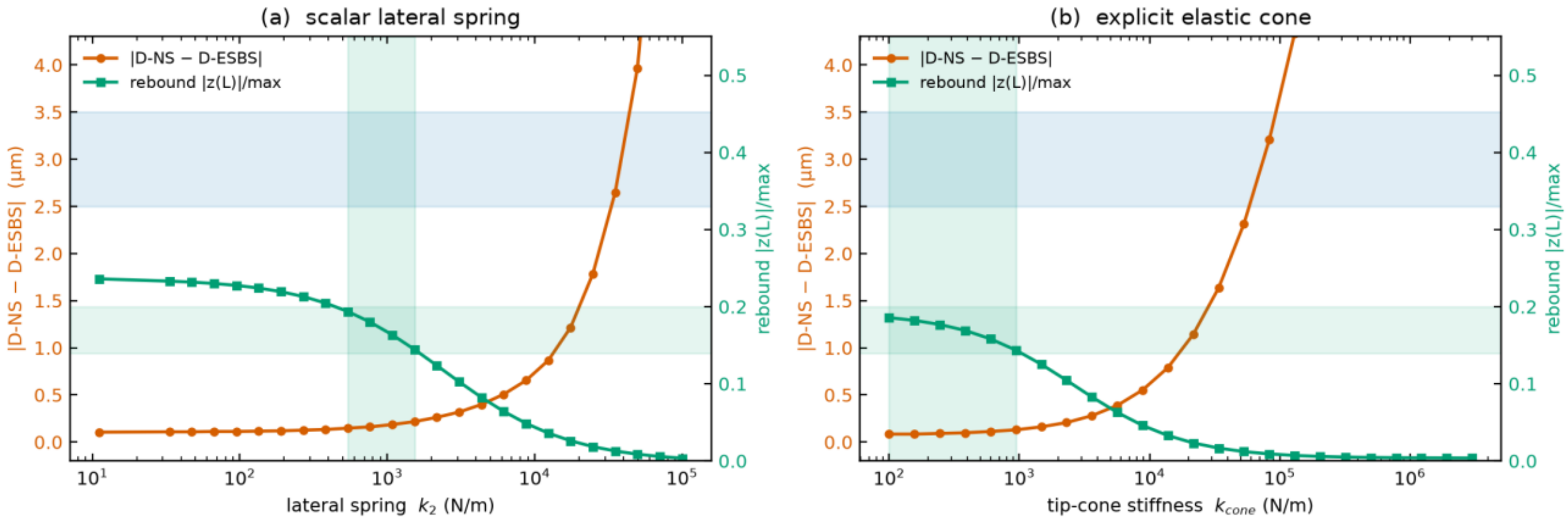

Figure S17: No admissible one-dimensional contact reproduces both the finite-element mode shape and the D-NS–D-ESBS separation. On-resonance near-tip rebound |z(L)|/max (green) and displacement-null separation |D-NS − D-ESBS| (orange) versus lateral stiffness at $k^* = 10^5$ N/m, for (a) a scalar lateral spring $k_2$ swept from the frictionless to the rigid limit and (b) an explicit elastic tip cone of stiffness $k_{cone}$. In both cases the rebound and the separation are anticorrelated: the stiffness window that reproduces the finite-element rebound (green band, 0.14–0.20; shaded) is disjoint from any that reaches the finite-element separation (blue band, 2.5–3.5 µm). Where the mode shape is correct the one-dimensional separation is only 0.03–0.2 µm; the ≈3 µm separation is reached only for a near-rigid contact that has lost the rebound. The result is identical at $k^* = 10^6$ N/m.